\documentclass[useAMS,usenatbib]{mn2e}

\usepackage{amssymb}
\usepackage{amsmath}
\usepackage{graphics}
\usepackage{graphicx}
\usepackage{epsfig,epstopdf,color,times}

\newcommand{\be}{\begin{equation}}
\newcommand{\ee}{\end{equation}}
\newcommand{\bea}{\begin{eqnarray}}
\newcommand{\eea}{\end{eqnarray}}
\def\ba#1\ea{\begin{align}#1\end{align}}

\newcommand{\refeq}[1]{Eq.~(\ref{eq:#1})}          
          
\newcommand{\reffig}[1]{Fig.~\ref{fig:#1}}          
          
\newcommand{\refsec}[1]{Sec.~\ref{sec:#1}}

\newcommand{\vs}{\nonumber\\} 

\def\bfr{\mathbf{r}}
\def\bfx{\mathbf{x}}
\def\bfv{\mathbf{v}}
\def\bfs{\mathbf{s}}
\def\bfk{\mathbf{k}}

\def\bfu{\mathbf{u}}

\def\rhat{\hat{\mathbf{r}}}
\def\nhat{\hat{\mathbf{n}}}
\def\khat{\hat{\mathbf{k}}}

\def\d{\mathrm{d}}

\def\Legendre#1#2{\mathcal{P}_{#1}(#2)}

\definecolor{royalblue}{rgb}{.25,.41,.88}

\title[Redshift-space 2PCF and baryon acoustic oscillations]
{The redshift-space galaxy two-point correlation function and
baryon acoustic oscillations} 

\author[Jeong et al.]
{Donghui Jeong$^{1,2,3}$, Liang Dai$^1$, Marc Kamionkowski$^1$, and
Alexander S. Szalay$^1$\\ 
${}^1$
Department of Physics and Astronomy,
Johns Hopkins University, 3400 N.\ Charles St., Baltimore, MD
21218 USA\\
${}^2$
Department of Astronomy and Astrophysics, 
The Pennsylvania State University, State College, PA 16802 USA\\
${}^3$
Institute for Gravitation and the Cosmos,
The Pennsylvania State University, State College, PA 16802 USA
}
\begin{document}
\twocolumn
\maketitle

\begin{abstract}
 Future galaxy surveys will measure baryon acoustic oscillations (BAOs) 
 with high significance, and a complete understanding of the
 anisotropies of BAOs in redshift space will be important to exploit the 
 cosmological information in BAOs. Here we describe the
 anisotropies that arise in the redshift-space galaxy two-point
 correlation function (2PCF) and elucidate the origin of features
 that arise in the dependence of the BAOs on the angle between
 the orientation of the galaxy pair and the line of sight. 
 We do so with a derivation of the configuration-space 2PCF using streaming 
 model. We find that, contrary to common belief, the
 locations of BAO peaks in the redshift-space 2PCF are anisotropic
 even in the linear theory. Anisotropies in BAO depend strongly
 on the method of extracting the peak, showing maximum $3$\% angular variation.
 We also find that extracting the BAO peak of $r^2\xi(r,\mu)$
 significantly reduces the anisotropy to sub-percent level angular variation.
 When subtracting the tilt due to 
 the broadband behavior of the 2PCF, the BAO bump is enhanced
 along the line of sight because of local infall velocities
 toward the BAO bump.  Precise measurement of the angular dependence
 of the redshift-space 2PCF will allow new geometrical tests of
 dark energy beyond the BAO.
\end{abstract}

\section{Introduction}
\label{sec:intro}
The galaxy two-point correlation function (2PCF) has been an
indispensable tool in modern physical cosmology.   Along with
the angular power spectrum of the cosmic microwave
background (CMB) temperature and polarization fluctuations
\citep{WMAP9:2013,Planck:2013}, luminosity distance measured
from Type-Ia supernovae
\citep{Conley/etal:2011,Suzuki/etal:2012}, and the local Hubble
parameter \citep{Riess/etal:2011,Freedman/etal:2012}, the
amplitude and shape of the galaxy 2PCF measured from large-scale
galaxy surveys such as the Sloan Digitial Sky Survey (SDSS)
\citep{anderson/etal:2014}, WiggleZ \citep{blake/etal:2011a},
and VIPERS \citep{delaTorre/etal:2013} have provided essential
constraints to the current $\Lambda$CDM cosmological model.
Measurements of the large-scale distribution of galaxies will 
continue with future surveys, such as
HETDEX\footnote{http://www.hetdex.org},
eBOSS\footnote{https://www.sdss3.org/future/eboss.php},
MS-DESI\footnote{http://desi.lbl.gov},
WFIRST-AFTA\footnote{http://wfirst.gsfc.nasa.gov} 
and
Euclid\footnote{http://sci.esa.int/euclid/} to mention a few, in
a deeper and wider manner, and the 2PCF will continue to be the
key observable in these surveys.

One of the main goals of these galaxy surveys is to constrain
the properties of dark energy using the baryon acoustic
oscillations (BAOs) in the galaxy 2PCF as a standard ruler (see,
e.g., \citet{DEreview}, for a recent review).
These BAOs appear as a bump in the galaxy 2PCF at galaxy
separations near $d_{\rm BAO}\simeq105~{\rm Mpc}~h^{-1}$, the
comoving distance that a baryon-photon acoustic wave
travels from the big bang to the baryon-decoupling epoch.  This
distance is now determined precisely by CMB measurements
\citep{WMAP9:2013,Planck:2013}.  From the observed galaxy 2PCF,
we can estimate the angular separation and redshift difference
corresponding to $d_{\rm BAO}$ which are translated to
measurements of, respectively, the angular-diameter distance
$D_A(z)$ and the Hubble expansion rate $H(z)$ at the observed
redshift $z$. These measurements constrain the energy density and
the equation of state of dark energy
\cite{seo/eisenstein:2007,Pritchard:2006ng}.

Since the first convincing detection of BAOs from the galaxy
power spectrum \citep{cole/etal:2005} and two-point correlation
function \citep{eisenstein/etal:2005}, comparison of theoretical
models \citep{seo/eisenstein:2007,seo/etal:2010} of BAOs
with measurements have yielded ever 
tighter constraints to dark-energy properties \citep{percival/etal:2007,
martinez/etal:2009,
kazin/etal:2010aa,
percival/etal:2010,
blake/etal:2011b,
blake/etal:2011a,
beutler/etal:2011,
anderson/etal:2012,
padmanabhan/etal:2012,
xu/etal:2012,
slosar/etal:2013,busca/etal:2013,
anderson/etal:2014}.

Thus far, most BAO analyses have considered only the angle averaged
2PCF (the ``monopole''), because the BAO bump itself is quite
subtle and has been hard to detect with high signal to noise.
The most recent analysis of the BOSS project in SDSS3 DR10 and
DR11 has reported a $7\sigma$ detection of the BAO bump in the
monopole 2PCF from a $13~{\rm Gpc}^3$ volume
\citep{anderson/etal:2014}.  For this level of signal to noise,
it is better to focus on robust statistics such as the monopole
and perhaps the quadrupole of the redshift-space galaxy 2PCF 
or the clustering wedges as discussed in, respectively,
\citet{padmanabhan/white:2008} and \citet{kazin/etal:2012}. 
The quadrupole 2PCF and clustering wedges have been measured from real
data in \citet{kazin/etal:2013} and
\citet{anderson/etal:2012,anderson/etal:2014}.
Note that one can still measure $D_A(z)$ and $H(z)$
separately from the monopole and quadrupole, since the monopole
2PCF is sensitive to  $D_V\equiv
\left[cz(1+z)^2D_A^2(z)/H(z)\right]^{1/3}$ and the quadrupole to
$D_A(z)H(z)$. Clustering wedges also lead to separate measurements of 
$D_A(z)$ and $H(z)$.

On the other hand, future galaxy surveys will map a volume
larger by an order of magnitude ($\sim 100~{\rm Gpc}^3$) and
will detect BAOs with much higher significance
($\sim20\sigma$).  They will thus allow measurement of the BAO
bump as a function of the orientation, relative to the line of
sight, of the galaxy pairs being correlated, or alternatively as
a function of the parallel (to the line of sight) separation
$s_\parallel$ and perpendicular separation $s_\perp$ of the two
galaxies.  This will allow a more direct and efficient
separation of the the measurement of the angular-diameter
distance $D_A(z)$ and the Hubble expansion rate $H(z)$.

The anisotropy in the galaxy 2PCF is due to
redshift-space distortions,\footnote{Strictly speaking, this
statement is true only when the  selection function is
independent of the line-of-sight directional velocity. For
example, the clustering of the high-$z$ ($z\simeq 6$)
Lyman-alpha emitters  strongly depends on the line-of-sight
velocity gradient due to the selection function
\citep{Zheng/etal:2011}.  Inaccurate dust correction can also
induce anisotropies in the galaxy 2PCF
\citep{Fang/etal:2011}.}\ the systematic distortion in the
galaxy 2PCF due to the Doppler (peculiar velocity) component of
the observed redshift.
The dependence of the BAO peak in the linear-theory
redshift-space galaxy power spectrum is straightforward, as the
anisotropies are well separated from the scale ($k$) dependence
($\nhat$ is the line-of-sight directional unit vector),
\be
     P_{g,s}(\bfk) = \left[b_g +f(\khat\cdot\nhat)\right]^2
     P_L(k),
\ee
where $b_g$ is the linear galaxy-bias parameter, and $f\equiv
\d\ln D /\d \ln a$ is the logarithmic derivative of the
linear-theory growth factor $D(a)$ as a function of scale factor
$a(t)$.  

However, it is useful or advantageous in many cases to make
measurements in configuration space, and in configuration space,
the corresponding galaxy 2PCF, and the dependence of the BAO
location on the angle between the galaxy pair and the line of
sight, shows a far richer behavior.
The standard lore here is, similar to
the angle averaged case, that the location of the BAO peak is
isotropic (robust) and anisotropies affect only the
amplitude and the width of the bump.  This is, however,
incorrect even for linear
redshift-space distortions: the location of the BAO peak
\textit{is} anisotropic, and the peak locations for different
angles differ from the real-space peak location.  This can be
easily seen from the formula for  the linear redshift-space 2PCF
\citep{Hamilton:1992,Hamilton:1997zq},
\ba
\xi_{g,s}(s,\mu)
=&
\left(b_g^2 + \frac23 b_gf + \frac{f^2}{5}\right)\xi_0^0(x)\Legendre{0}{\mu}
\vs
&-
\left(\frac43b_gf + \frac47f^2\right)\xi_2^0(x)\Legendre{2}{\mu}
+\frac{8}{35}f^2 \xi_4^0(x)\Legendre{4}{\mu}.
\label{eq:xigs}
\ea
In \refeq{xigs}, the radial dependence is given by the
integral of the spherical-Bessel-weighted linear matter power
spectrum $P_L(k)$,
\be
     \xi_{n}^m(x) \equiv \int \frac{k^2 \d k}{2\pi^2} P_L(k)
     \frac{j_n(kx)}{(kx)^m},
\label{eq:xinm}
\ee
where $j_l(x)$ is a spherical Bessel function,
and the angular dependence is given by the Legendre polynomial
$\Legendre{\ell}{\mu}$ with $\mu$ being the cosine of the angle
between the line of sight and the separation. 
\begin{figure}
\centering
\includegraphics[width=0.498\textwidth]{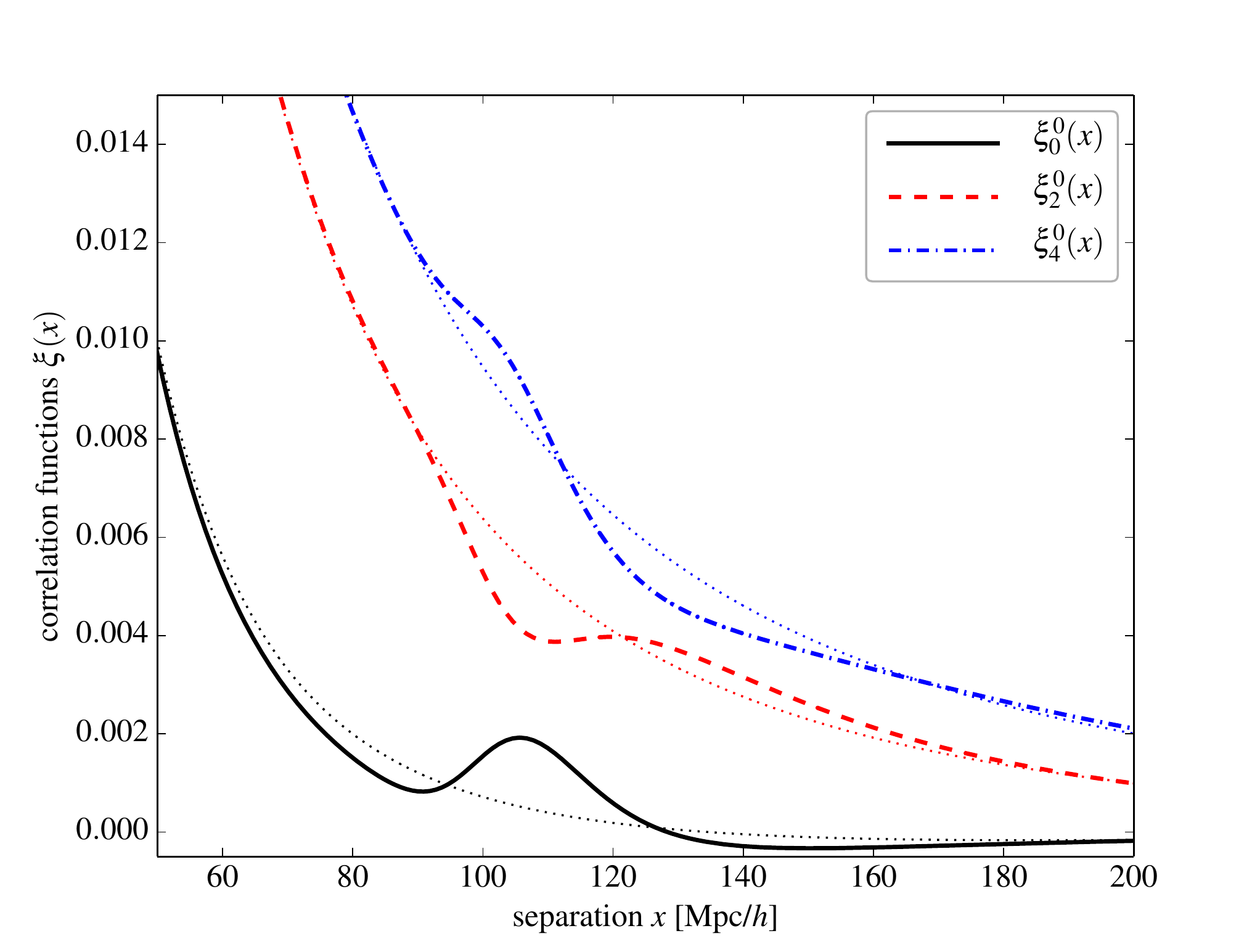}
\caption{
     The correlation functions $\xi_0^0(x)$ (black solid line),
     $\xi_2^0(x)$ (red dashed line), and $\xi_4^0(x)$ (blue
     dot-dashed line) defined in \refeq{xinm} at the present
     time ($z=0$) with galaxy bias $b_g = 1$.  The dotted curves
     show $\xi_n^m(x)$ with the no-wiggle power spectrum given
     in \citet{Eisenstein:1997ik} to highlight the BAO bump in
     each curve.
}
\label{fig:xis}
\end{figure}
We plot in Fig.~\ref{fig:xis} the three radial functions
$\xi_0^0(x)$,  $\xi_2^0(x)$, and $\xi_4^0(x)$.  All three radial
functions $\xi_n^m(x)$ show BAOs, but the BAO
bumps appear at different separations for the different functions. Therefore,
since the 2PCF is given for different orientations with respect
to the line of sight by the linear combination of three radial
functions in \refeq{xigs}, the BAO peak location in two
dimensions is anisotropic and varies with the line-of-sight angle.  
Examples of the anisotropies are shown in the bottom right panel
of \reffig{xi2d_streaming} and \reffig{xi2d_streaming_larger},
respectively, for the BAO scales and for the very large scales.
The bottom line is that the redshift-space distortion induces
anisotropies in the {\it location} of the BAO

The purpose of this paper is to develop a heuristic
understanding of these anisotropies.  The second goal is to
study geometrical tests of dark energy that use these
two-dimensional configuration-space BAO anisotropies.
Below we will explain in greater detail how the anisotropies are
generated from the velocity field. With this understanding,
we will be able to go beyond simply the usual distance
measurements from BAO, as the shape of angular contours of the
redshift-space 2PCF can be used for the broadband
Alcock-Paczynski test \citep{song/etal:2013} and for a dynamical
measure of dark energy by obtaining $f\sigma_8$
from \refeq{xigs} \citep{song/etal:2010,song/etal:2011,
beutler/etal:2012,reid/etal:2012,delaTorre/etal:2013}.
Note that the $\sigma_8$ dependence comes through the amplitude of the
linear matter power spectrum.  Moreover, features like the
zero crossing in the right-bottom corner of
\reffig{xi2d_streaming_larger} may provide a useful
probe for, e.g. matter-radiation equality \citep{prada/etal:2011}.

The rest of this paper is organized as follows.  In
\refsec{xigs_model}, we construct a heuristic but concrete
physical model for the redshift-space 2PCF in linear theory.  We
show that the terms proportional to $\beta$ in \refeq{xigs}
originate from the variation in the bulk motion of galaxies
while the $\beta^2$ terms account for the variation in the
pairwise velocity dispersion.  Based on this model, we then
discuss the anisotropies in the BAO bump in greater detail in
\refsec{BAO}.  We conclude with a discussion of possible
future research directions in \refsec{conclusion}.

\section{Anisotropies in the redshift-space 2PCF}
\label{sec:xigs_model}
In this Section, we present a physical explanation for the
anisotropies in the galaxy 2PCF.
First, in \refsec{xi_anisotropy_heuristic}, we construct from
the linear growth rate and the continuity equation the
peculiar-velocity field around a \textit{typical} galaxy.
We then calculate the redshift-space 2PCF by including
the line-of-sight directional component of the peculiar velocity.
Although this captures the dominant effect, this heuristic argument 
must be complemented by the spatial variation of the
second-order moment of the velocity field in order to completely
account for the the linear redshift-space distortion.
In \refsec{xi_anisotropy_streaming}, we demonstrate this with the
streaming model
\citep{Peebles:1980book,Fisher:1993ye,Fisher:1994ks,Scoccimarro:2004tg}
that describes the redshift-space 2PCF by convolution between
the real-space 2PCF and the pairwise-velocity distribution function.

\subsection{Understanding anisotropies: a heuristic argument}
\label{sec:xi_anisotropy_heuristic}
\begin{figure*}
\centering
\includegraphics[width=0.498\textwidth]{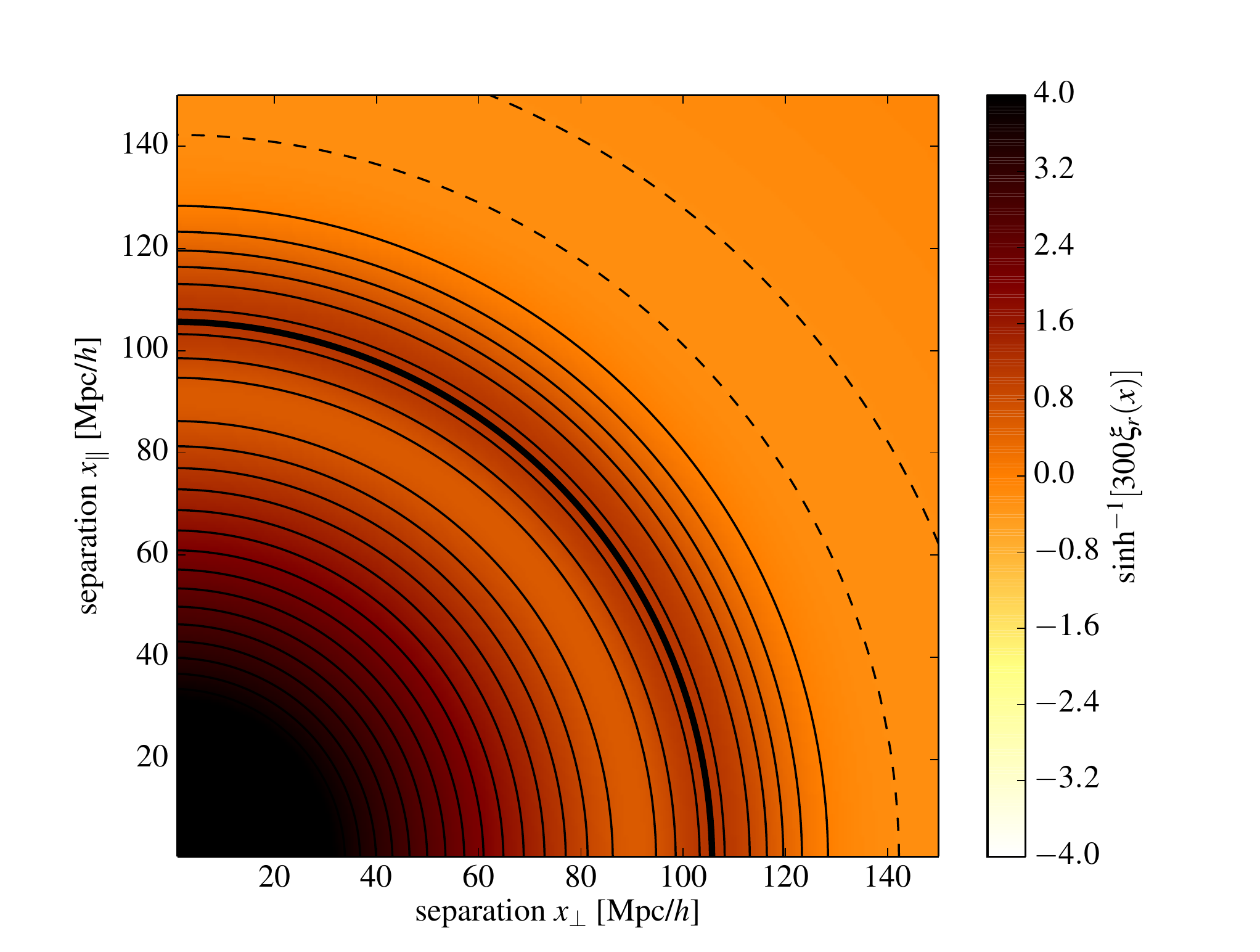}
\includegraphics[width=0.498\textwidth]{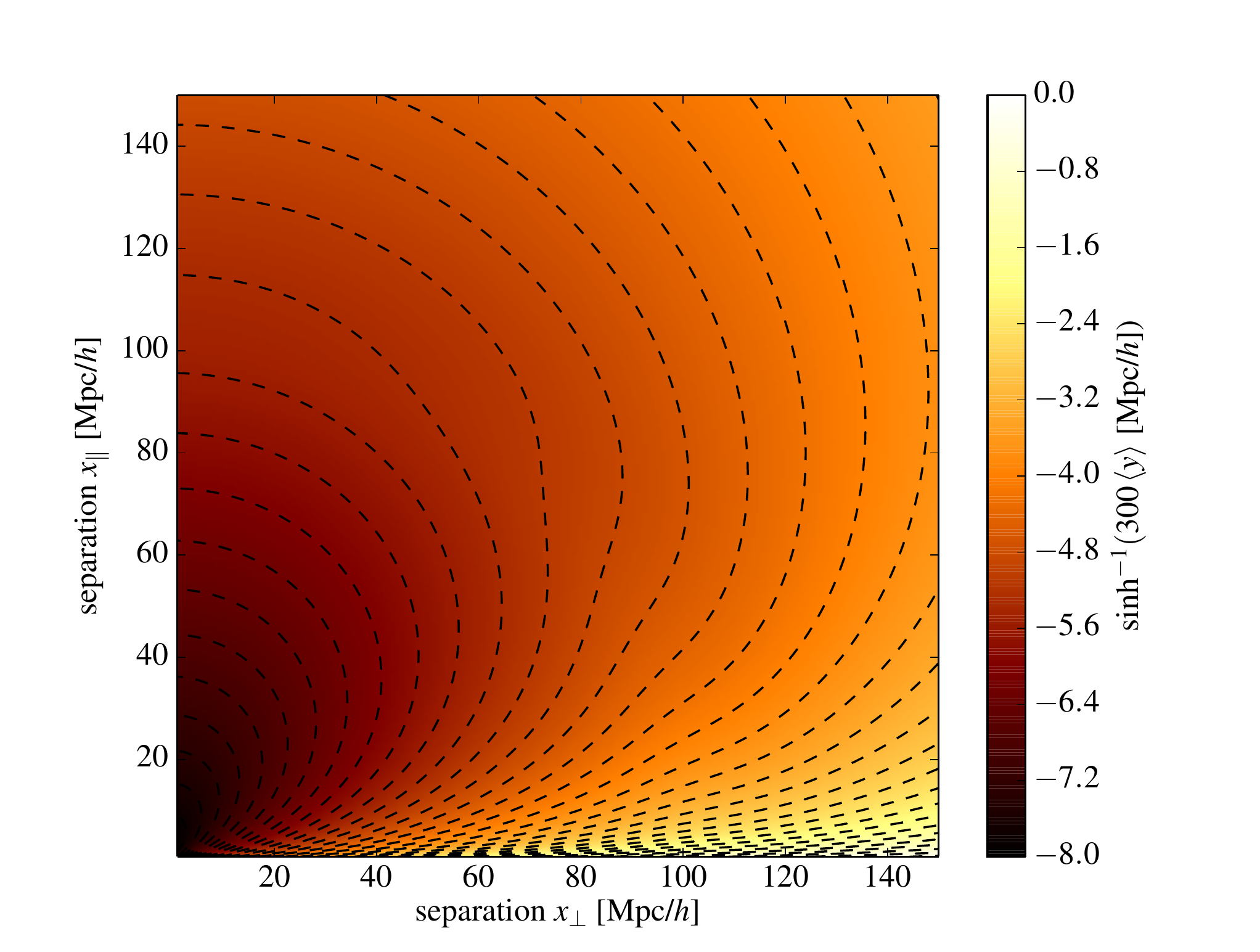}
\includegraphics[width=0.498\textwidth]{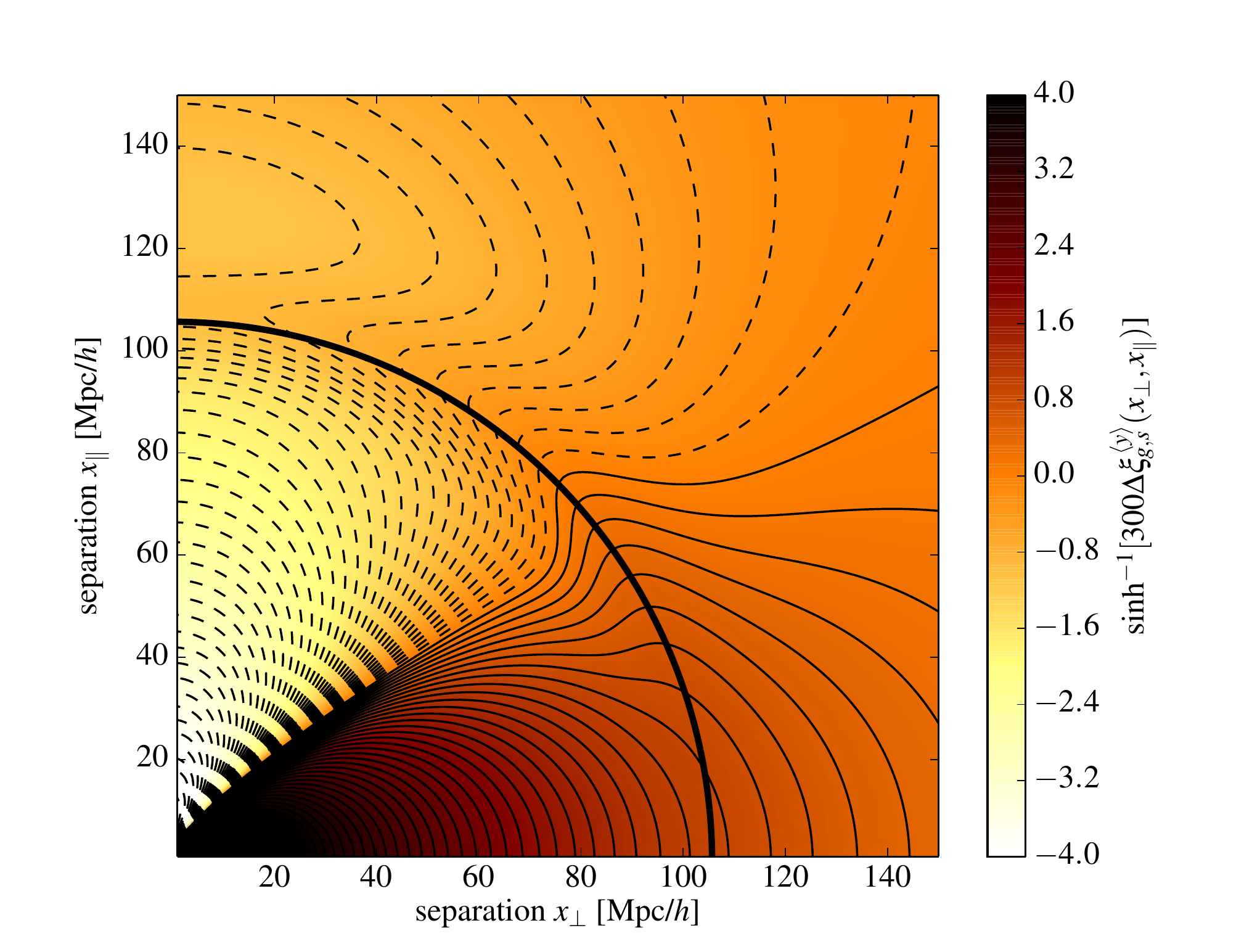}
\includegraphics[width=0.498\textwidth]{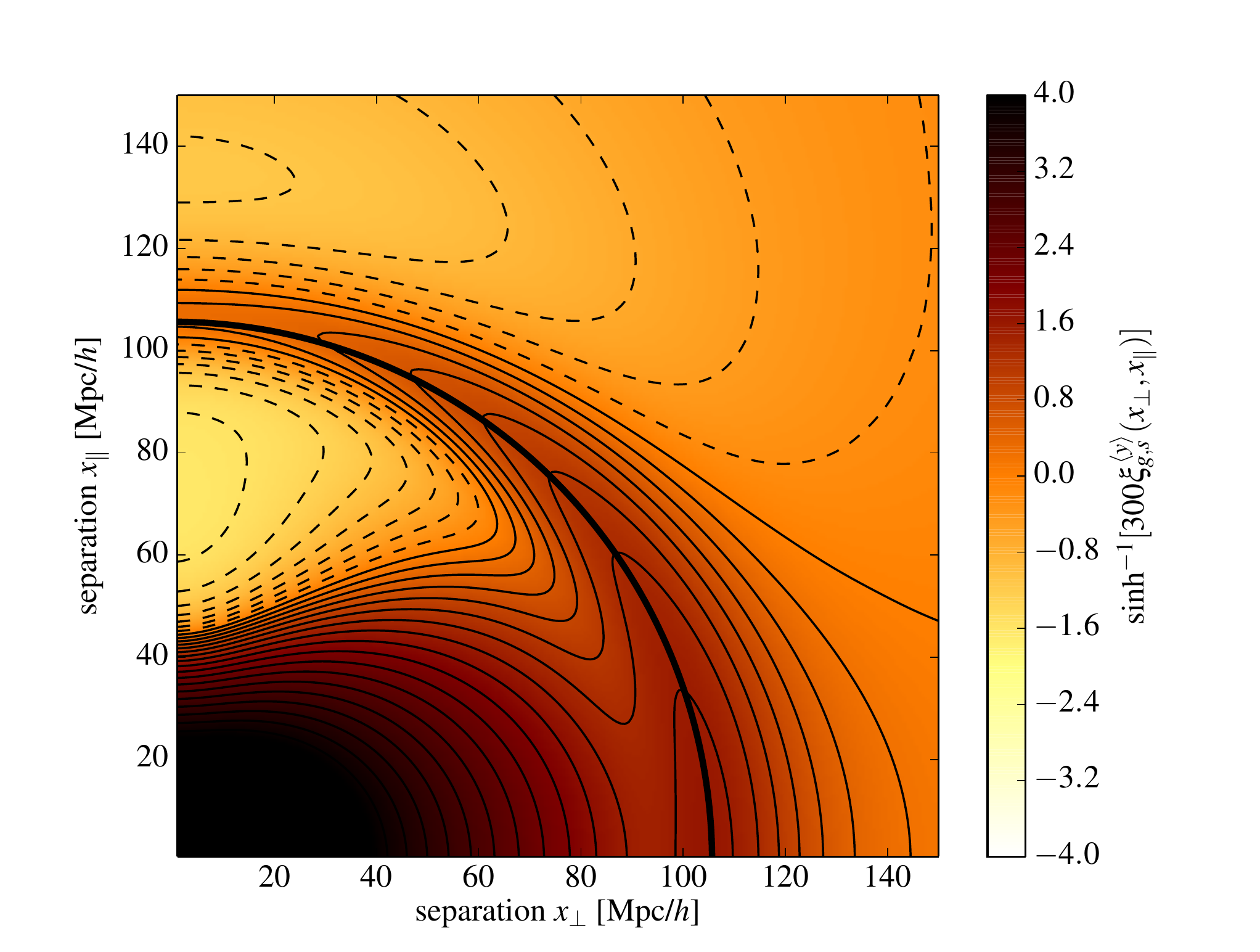}
\caption{
The two-dimensional $(x_\perp,x_\parallel)$ 2PCFs from the heuristic 
argument given in \refsec{xi_anisotropy_heuristic}.
These plots are for $b_g=2$ galaxies at $z=0.5$.
For each plot, we show the contour 
lines for arc-sine-hyperbolic of the function as indicated in the label
next to the colorbar. The value exceeding the colorbar range is painted
as the corresponding extreme colors.
Contour lines are drawn for integer multiples of $1/6$ 
($1/9$ for the $\Delta \xi_{g,s}^{\left<y\right>}$),
and dashed lines are for negative values.
\textit{Top left:}
Two-dimensional contour of the real-space 2PCF. Without redshift-space
distortions, the 2PCF is isotropic. We can think of the real-space 2PCF as a
typical density profile around a galaxy. One can see the BAO bump around 
the isotropic ring with $x \approx 106~\mathrm{Mpc}/h$.
\textit{Top right:}
The mean displacement $\left< y\right>$ along the LoS direction.
The displacement vector is along the negative-$x_\parallel$ direction, 
and brighter (darker) colors show lower (higher) amplitudes.
For a given separation, the line-of-sight direction shows larger 
displacement, while the displacement field itself is a decreasing function of 
separation.
\textit{Bottom left:}
Correction to the apparent density profile (2PCF) due to the velocity field. 
The correction is given by the LoS derivative of the the 
line-of-sight displacement shown in the \textit{Top right} panel. 
While the projection effect enhances the apparent clustering toward the 
perpendicular direction, the reduction of the displacement amplitude reduces 
the apparent clustering toward the parallel direction.
\textit{Bottom right:}
Final two-dimensional density distribution (2PCF) from the heuristic argument 
combining the real space isotropic density distribution (2PCF) and
anisotropic correction due to the scale-dependent LoS displacement.
}
\label{fig:xi2d_heuristic}
\end{figure*}
Consider a spherically symmetric distribution of galaxies 
with a radial profile of number density $n_g(x,t) = \bar{n}_g(t)
\left[1+\xi_g(x,t)\right]$, where
$\xi_g(x)$ is the real-space galaxy 2PCF,
\be
\xi_g(x,t) 
=
\left<\delta_g(\bfx_1,t)\delta_{g}(\bfx_0,t)\right>,
\ee 
where $\delta_g$ is the fractional galaxy density perturbation.
Here, we denote the location of the central galaxy by $\bfx_0$, and 
other galaxies by $\bfx_1$ so that $x = |\bfx_1-\bfx_0|$.
As the galaxy 2PCF measures the excess number of neighboring
galaxies as a function of distance, the quantity
$\bar{n}_g(1+\xi_g)$ provides the mean radial profile of the
galaxy density around a \textit{typical} galaxy in the Universe.

In the top left panel of \reffig{xi2d_heuristic} (see also the
black solid curve in \reffig{xis}), we show a contour plot of
the radial density profile around the central galaxy 
(linear theory galaxy 2PCF) as a function of the parallel
separation $x_\parallel$ and perpendicular separation
$x_\perp$ to the line of sight.
For smaller radii, the galaxy density decreases monotonically until it
reaches the BAO bump around $x\approx 105~\mathrm{Mpc}/h$ and
then decreases again after the BAO peak.  At around
$x\approx150~\mathrm{Mpc}/h$, the density profile reaches a
minimum and then increases afterwards to reach the cosmic mean in
the $x\to\infty$ limit.
We show the contour plot for $\sinh^{-1}(300\xi_g)$,
which scales linearly when $|\xi_g|<1/300$ but logarithmically for larger
arguments, to represent the large dynamical range of the 2PCF.

In linear theory, the galaxy density contrast evolves in time as 
$\delta_g \equiv n_g(x,t)/\bar{n}_g(t)-1  = \xi_g(x,t) \propto D^2(t)$, with 
the linear growth factor $D(t)$. Then, from the linear bias relation, 
we estimate the matter density contrast as $\delta_m = \delta_g/b_g$
with the linear bias parameter $b_g$.
Once the time evolution of the matter density contrast is known,
the linearized continuity equation,
\ba
&\frac{\d\delta_m(\bfx,t)}{\d t}
+
\frac{1}{a}\nabla_\bfx \bfu(\bfx,t) = 0,
\label{eq:continuity}
\ea
allows us to calculate the corresponding peculiar-velocity field,
\ba
\bfu(\bfx,t) 
=
-2 \frac{aHf}{b_g} 
\nabla_{\bfx}
\left[
\nabla_{\bfx}^{-2}\xi_g(\bfx,t)
\right].
\label{eq:heuristic_continuity}
\ea
Here, $\nabla_{\bfx}^{-2}$ is the inverse Laplacian, and we use 
$dD^2/dt = 2 aHfD^2$ with the scale factor $a$, the 
Hubble expansion rate $H$, and logarithmic growth factor 
$f\equiv \d\ln D/\d\ln a$.
The right-hand side of \refeq{heuristic_continuity} can be
written as an integral, by using the Green's function
$|\bfx-\bfx'|^{-2}$ of the three-dimensional Laplacian, as
\be
\bfu(\bfx,t) 
=
-2 \frac{aHf}{b_g} 
\frac{\bfx}{x^3} \int_0^x dy \xi_g(y) y^2.
\ee
To facilitate comparison to \refeq{xigs},
we use the $\ell=1$ case of the identity [Eq. (20) in
\citep{Matsubara:1996nf}],
\ba
\xi_{2\ell}^0(x_0)
=
\frac{(-1)^\ell}{x_0^{2\ell+1}}
\left(
\prod_{i=1}^{\ell}
\int_0^{x_{i-1}} x_idx_i
\right)
x_{\ell}^{2\ell}
\left(
\frac{d}{dx_{\ell}}\frac{1}{x_{\ell}}
\right)^\ell x_{\ell}\xi_0^0(x_\ell),
\ea
to rewrite the velocity field as 
\be
\bfu(\bfx,t) 
=
-\frac23 aHfb_g 
\left[
\xi_0^0(x) + \xi_2^0(x)
\right]
\bfx
\label{eq:heuristic_velocity},
\ee
where $\xi_n^m(x)$ is defined in \refeq{xinm}.
As we are considering a spherically symmetric distribution of galaxies,
the velocity field $\bfu(\bfx,t)$ is purely radial and aimed at
the central-galaxy location $\bfx_0$.

When we look at this spherical distribution of galaxies in
redshift space (with line-of-sight direction $\nhat$), 
galaxies at $\bfx$ appear to be at $\bfx \to \bfs=\bfx +
y\nhat$, where 
\be
y 
= \frac{\bfu\cdot\nhat}{aH} 
= 
-\frac23 fb_g 
x_\parallel
\left[
\xi_0^0(x) + \xi_2^0(x)
\right],
\ee
is the displacement due to the line-of-sight component of the
peculiar velocity $\bfu$.  We show the displacement $y$ in the
top right panel of \reffig{xi2d_heuristic}.  Two competing
effects determine the anisotropies in the displacement 
$y$: the amplitude of the velocity and the line-of-sight projection.
For a fixed line-of-sight angle, the amplitude of the displacement 
is a decreasing function of $x$. For a fixed separation $x$, 
the line-of-sight projection of a purely radial velocity field diminishes
the displacement along the perpendicular direction and 
maximizes it along the parallel direction.
Adding up the two effects, the displacement field is smaller along the
perpendicular direction and for large separations, and it has a
maximum amplitude ($y$ is negative as the velocity is toward the
center) along the parallel direction.

Since the total galaxy number stays the same in real and redshift
space, we find the redshift-space density contrast,
\ba
\delta_{g,s}(\bfx) 
=&
\left(1 + \delta_g(\bfx) \right)\left|\frac{\partial^3 s}{\partial^3 x}\right|^{-1}-1
\simeq
\delta_{g}(\bfx)
-
\frac{\d y}{\d x_\parallel}.
\ea
The leading-order redshift-space distortion in peculiar velocity
is therefore given by the line-of-sight derivative of the
displacement, which itself is the line-of-sight projection of the 
peculiar velocity. This can be readily understood because constant 
line-of-sight velocity would only shift all galaxies by the same amount, 
and the density contrast only shifts up and down between real and redshift 
space. We work out the linear correction,
\ba
\frac{\d y}{\d x_\parallel}
=&
-\frac23 fb_g\left[\xi_0^0(x)+\xi_2^0(x)\right]
+ 2 fb_g\mu^2\xi_{2}^0(x),
\label{eq:yterm}
\ea
because $\partial x/\partial x_{\parallel} = x_\parallel/x = \mu$,
and 
\be
\frac{\d}{\d x}\left[\xi_0^0(x) + \xi_2^0(x)\right]
=-\frac{3}{x}\xi_2^0(x)
\ee
from the identities,
\ba
\frac{\d j_\ell(x)}{\d x}
=&
\frac{\ell}{x}j_{\ell}(x) - j_{\ell+1}(x)
\vs
j_{\ell-1}(x) + j_{\ell+1}(x) 
=& 
\frac{2\ell+1}{x}j_{\ell}(x).
\ea
The $\mu^2$ term in \refeq{yterm} accounts for the anisotropic correction 
due to the inflow of galaxies toward the center, which suppresses the
density contrast along the parallel directions.  The resulting
redshift-space-density contrast is highly anisotropic: the
enhancement is along the perpendicular direction and suppression
along the parallel direction as shown in the bottom right panel
of \reffig{xi2d_heuristic}.  Note that the correction term due
to the bulk velocity (bottom-left panel) has a peculiar feature
around BAO scales (shown with the thick black line) that
anisotropically shifts the BAO peak location.  We discuss the
shift of the BAO peak more in \refsec{BAO}. 

Just as we construct the galaxy density distribution from the
real-space 2PCF, we interpret the anisotropic density
distribution in redshift space as the redshift-space 2PCF:
\ba
\xi_{g,s}(x,\mu)
=
\left[
b_g^2
+
\frac23
fb_g
\right]
\xi_0^0(x)
-
\frac43
fb_g
\xi_2^0(x)
\Legendre{2}{\mu},
\label{eq:heuristic_final}
\ea
with $\Legendre{2}{\mu} = \frac12 (3\mu^2-1)$ the Legendre polynomial.
The redshift-space 2PCF, \refeq{heuristic_final}, that
we have derived from the heuristic argument here correctly reproduces 
the terms proportional to $f$ in the usual expression in \refeq{xigs}.
This argument, based on the mean streaming velocity, cannot,
however, reproduce the full linear-order redshift-space distortion.  
We will need to include in the following Section the scale-dependent
velocity dispersion to reproduce the terms proportional to
$f^2$.

\subsection{The streaming model and linear redshift space distortion}
\label{sec:xi_anisotropy_streaming}
\begin{figure*}
\centering
\includegraphics[width=0.498\textwidth]{figs/plot_xis_yterm_2d.pdf}
\includegraphics[width=0.498\textwidth]{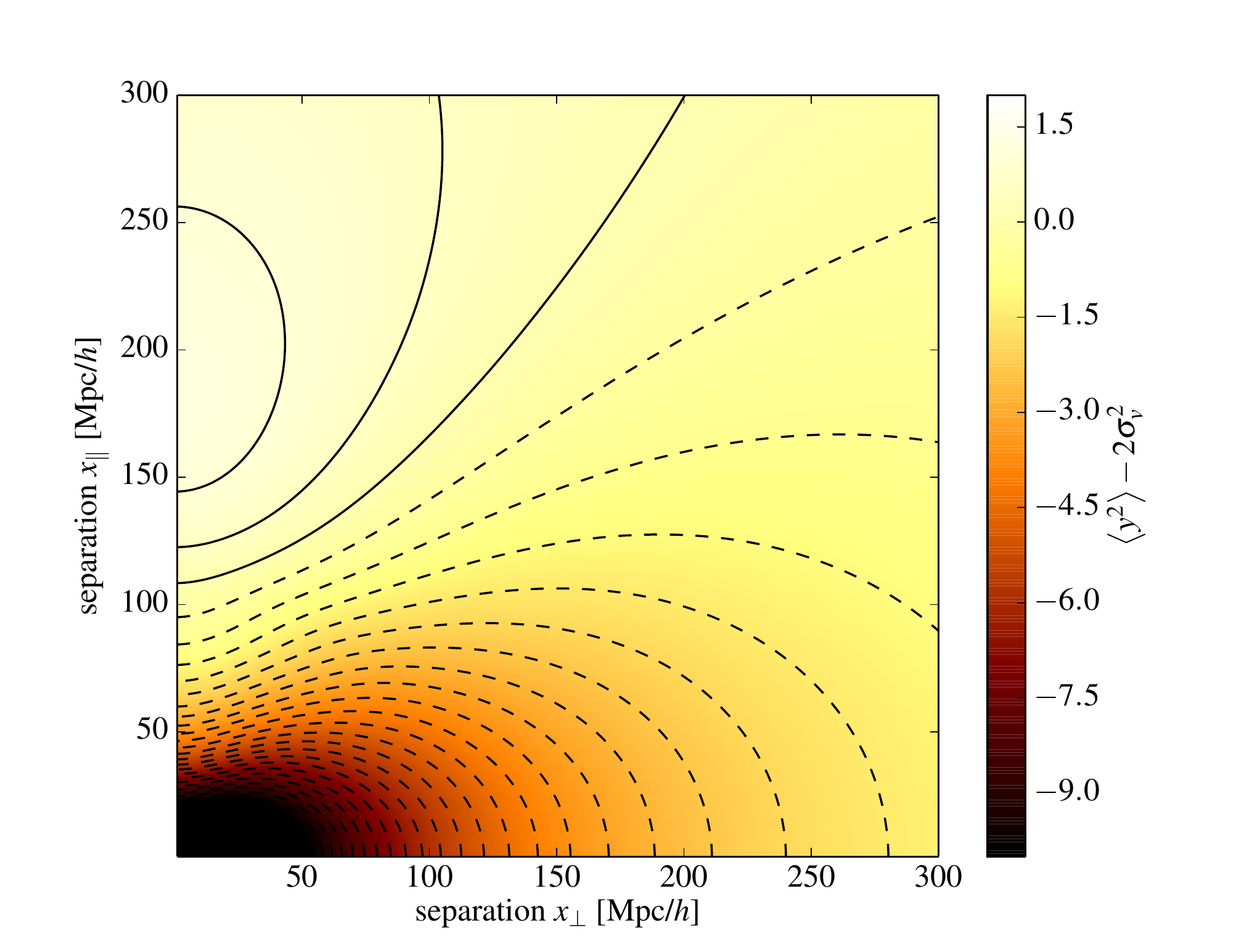}
\includegraphics[width=0.498\textwidth]{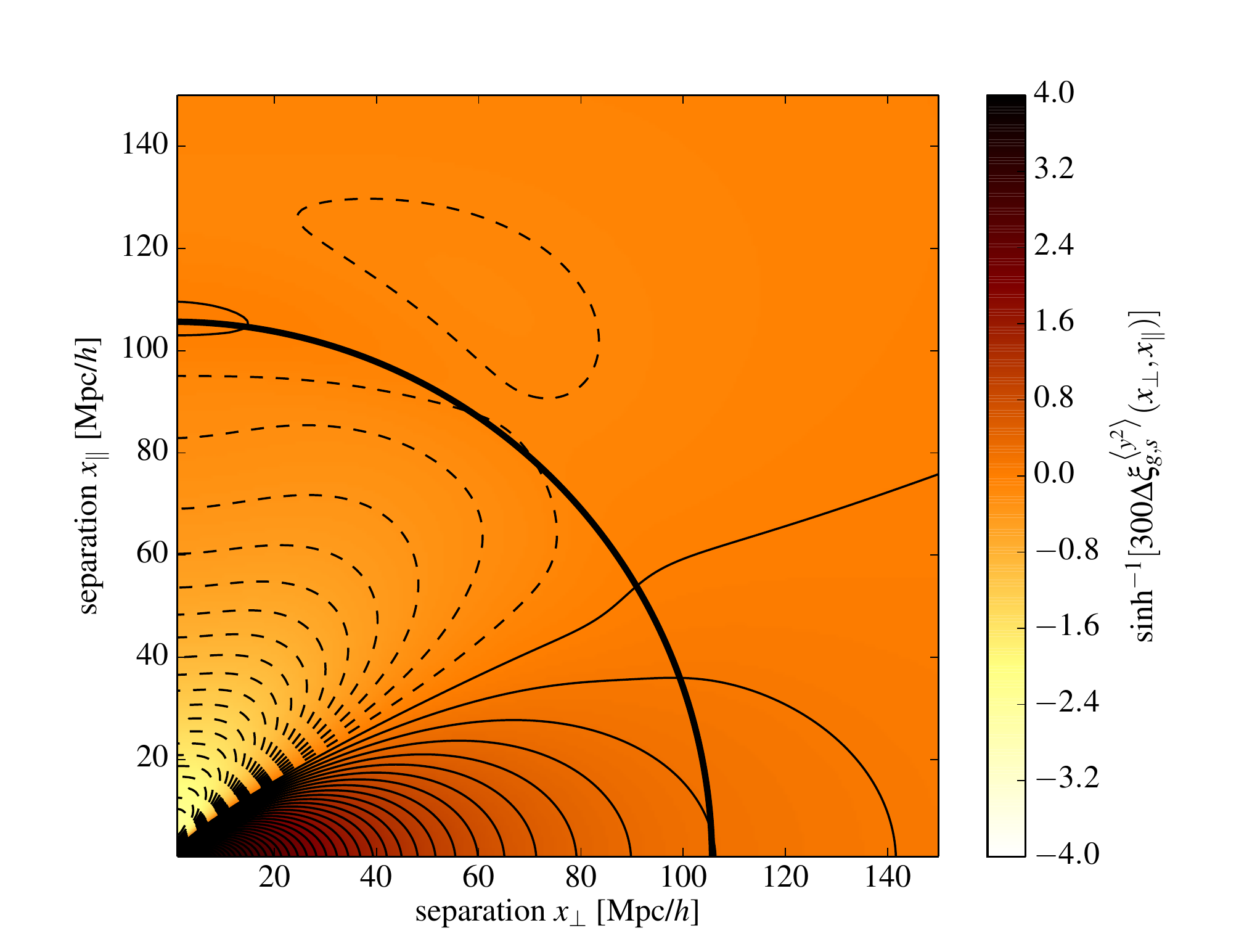}
\includegraphics[width=0.498\textwidth]{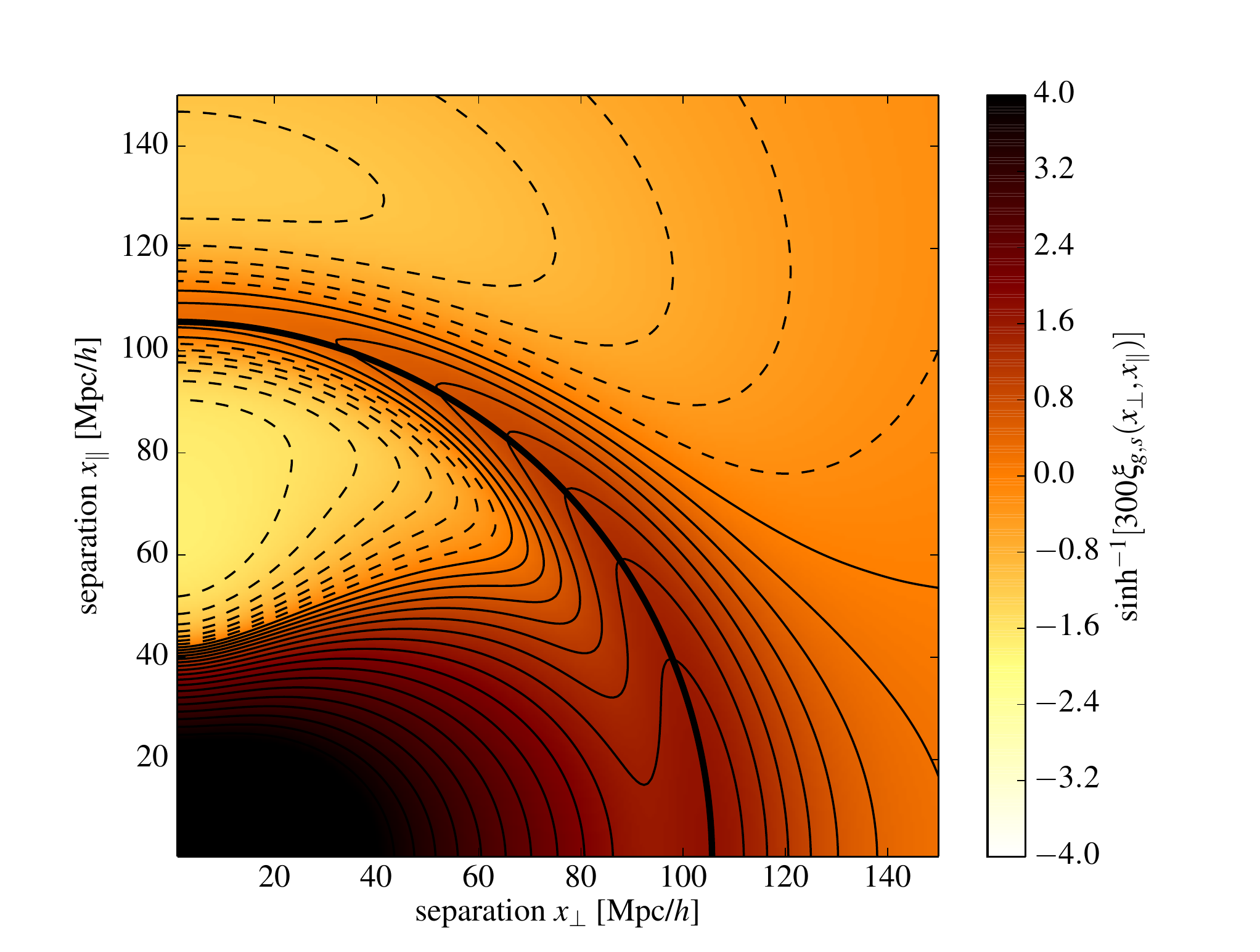}
\caption{
Two-dimensional $(x_\perp,x_\parallel)$ 2PCF from the streaming model.
The color scheme and contour lines are the same as \reffig{xi2d_heuristic} 
except for the top right panel where the color scale is linear and contour 
lines show multiples of half-integer.
Contour lines are for multiples of $1/9$ for
$\Delta \xi_{g,s}^{\left<y^2\right>}$.
\textit{Top left:} Redshift-space 2PCF including only the effect from 
mean displacement $\xi_g - d/dx_\parallel\left<y\right>$. This plot is 
the same as the bottom right panel of \reffig{xi2d_heuristic}.
\textit{Top right:}
The scale and angular dependence of the second-order moment 
$\left<y^2\right> - 2\sigma_v^2$ of the line-of-sight pairwise velocity 
field from linear theory. To show the angular dependence more clearly, 
we show a wider range of scales. The second moment $\left<y^2\right>$
asymptotes to the constant value $\left<y^2\right>\to 2\sigma_v^2$ for
large separations ($x\to\infty$). 
Mostly $\left<y^2\right> \lesssim 2\sigma_v^2$ except along the
line of sight and 
$x\gtrsim 100~\mathrm{Mpc}/h$ where $\xi_1^3(x) < \xi_2^2(x)$.
Along the parallel direction, $\left<y^2\right>$ reaches the maximum around 
$x\approx 180~\mathrm{Mpc}/h$, and decreases to $2\sigma_v^2$.
The correction to the 2PCF is given by the second derivative of
the second moment. 
\textit{Bottom left:}
Correction to the redshift-space 2PCF from the second momentum of 
the displacement 
$\Delta\xi_{g,s}^{\left<y^2\right>}\equiv \frac12 
d^2/dx_\parallel^2\left<y^2\right>$.
Like the contribution from the mean displacement (the bottom left 
panel of \reffig{xi2d_heuristic}), the anisotropic correction
shows clear trend:
redshift-space distortion decreases (negative correction) 
the correlation function along the LoS direction
and increases (positive correction) along the direction perpendicular to
the line of sight.  However, the amplitude of the anisotropy is
sub-dominant to the effect from the mean displacement.
\textit{Bottom right:}
Redshift-space 2PCF from linear Kaiser expression \refeq{xigs}.
The contribution from the second order moments deepens the negative valley 
along the line-of-sight direction and increases the perpendicular directional
clustering amplitude. Overall, however, the anisotropies are dominated by
the mean displacement.
}
\label{fig:xi2d_streaming}
\end{figure*}
\begin{figure*}
\centering
\includegraphics[width=0.498\textwidth]{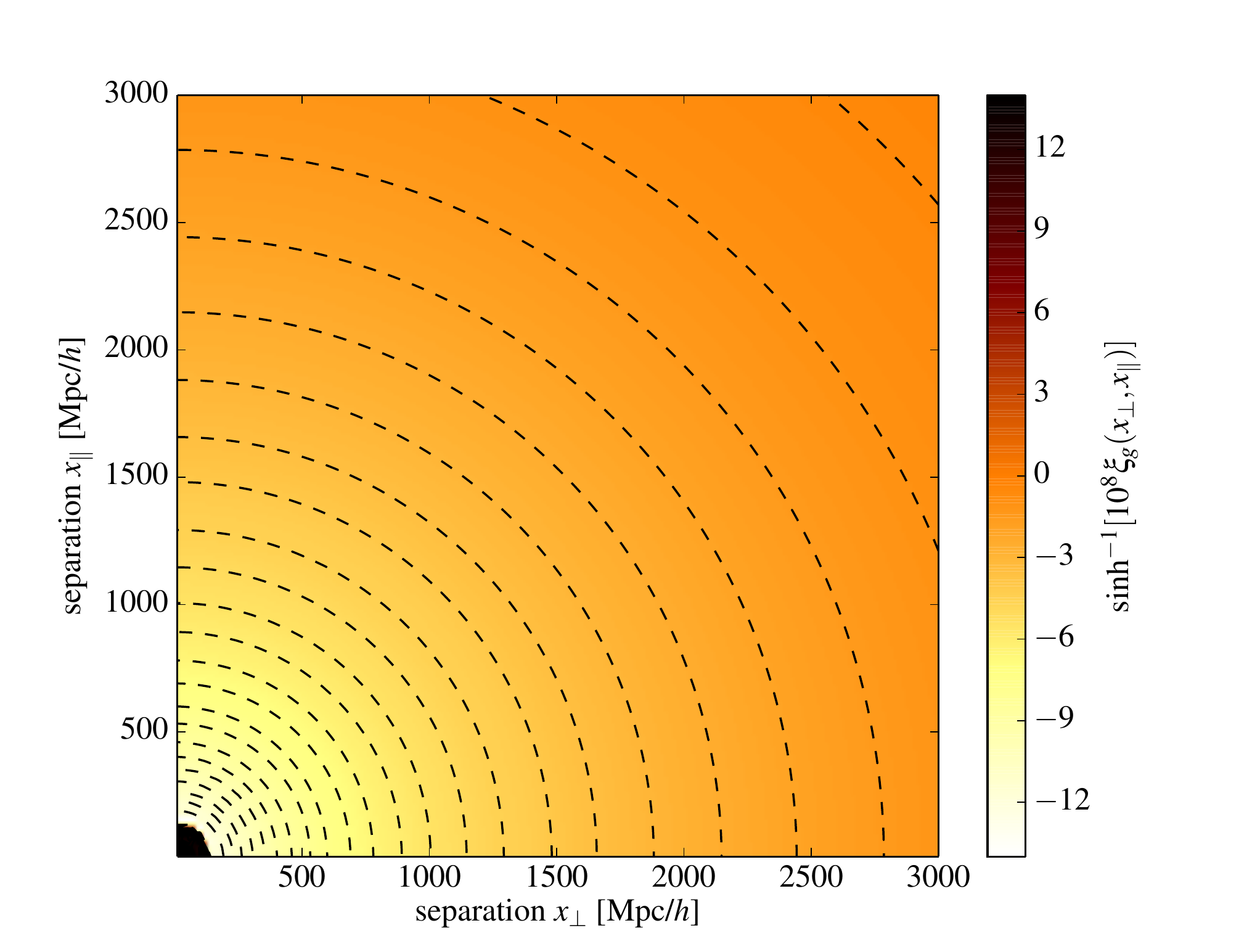}
\includegraphics[width=0.498\textwidth]{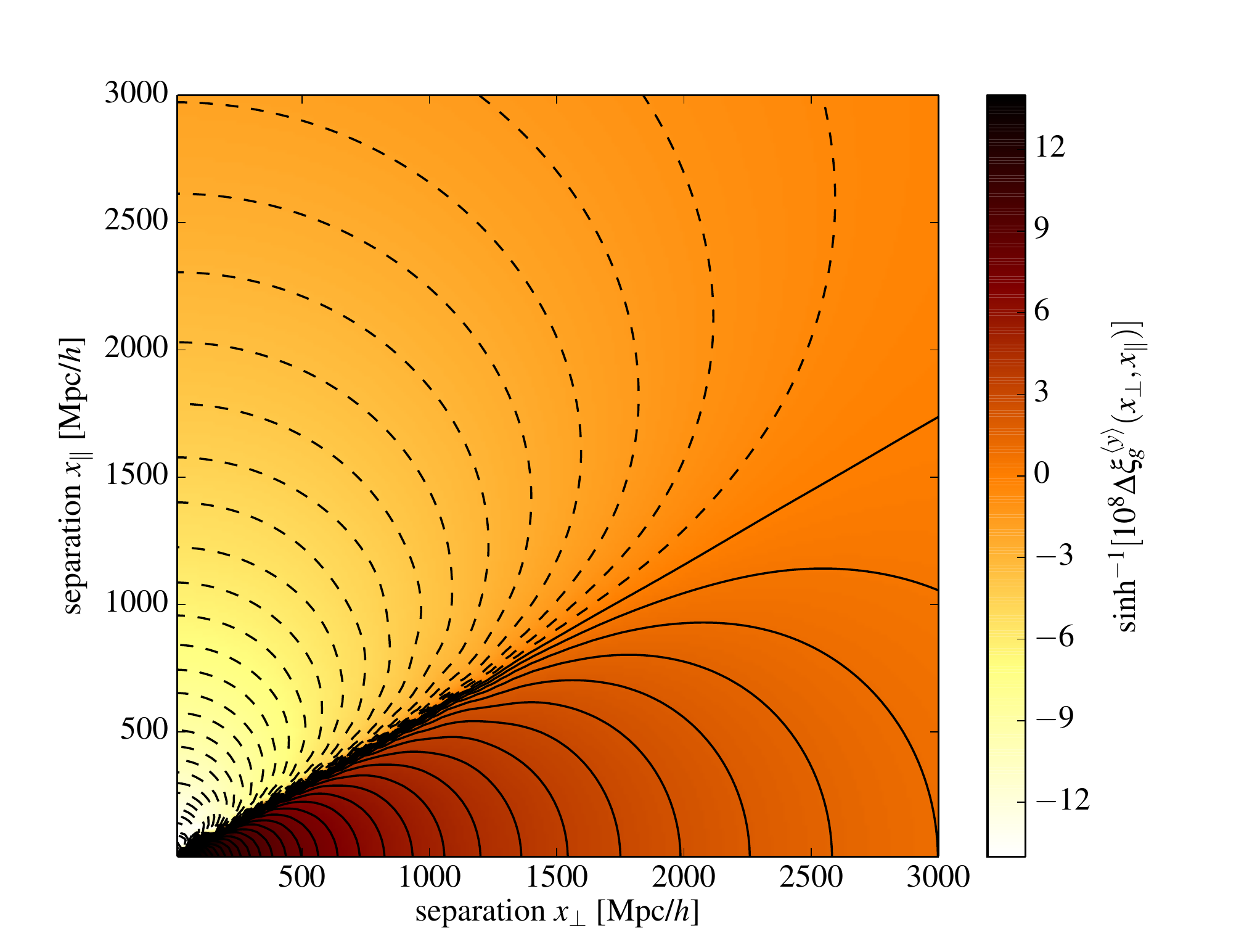}
\includegraphics[width=0.498\textwidth]{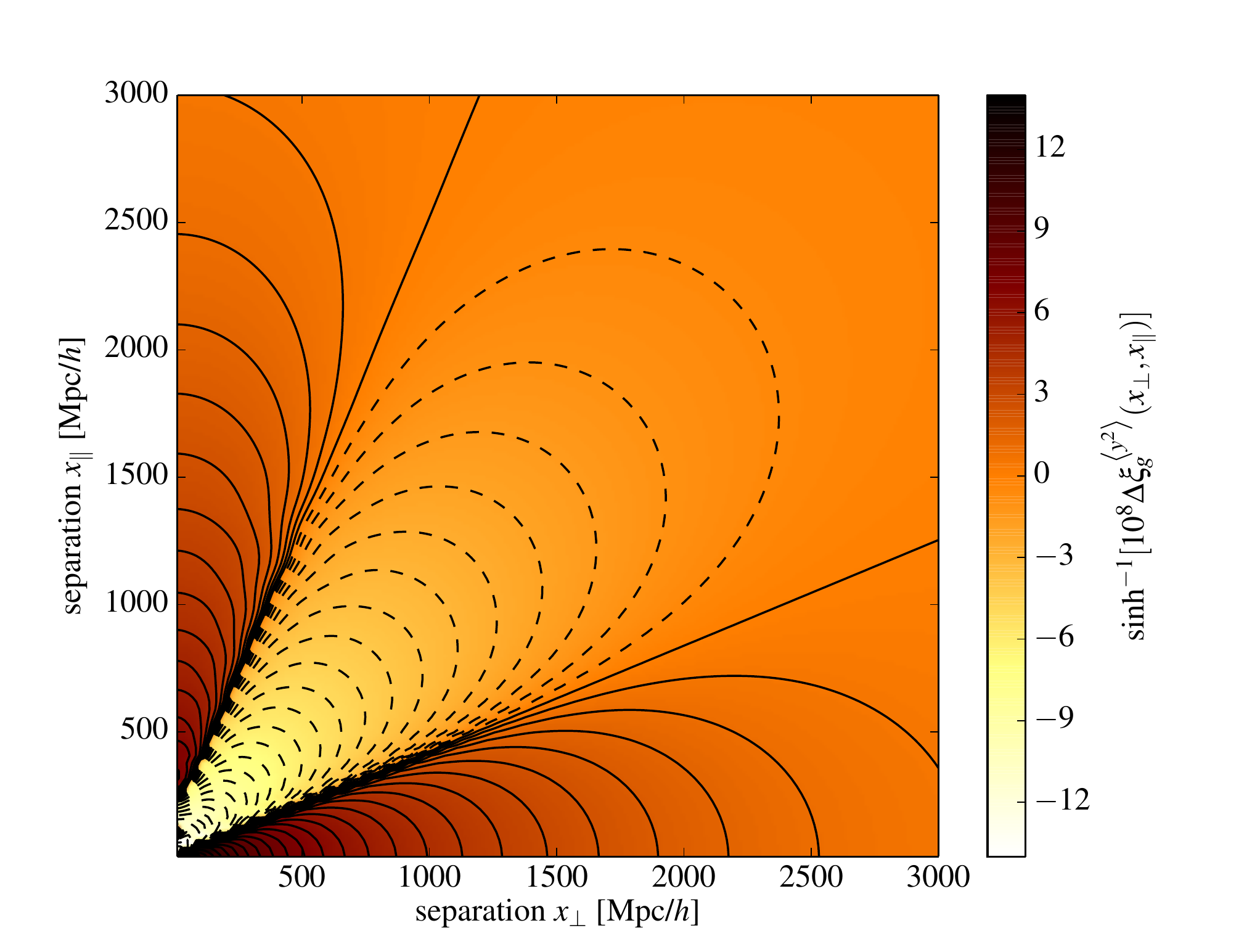}
\includegraphics[width=0.498\textwidth]{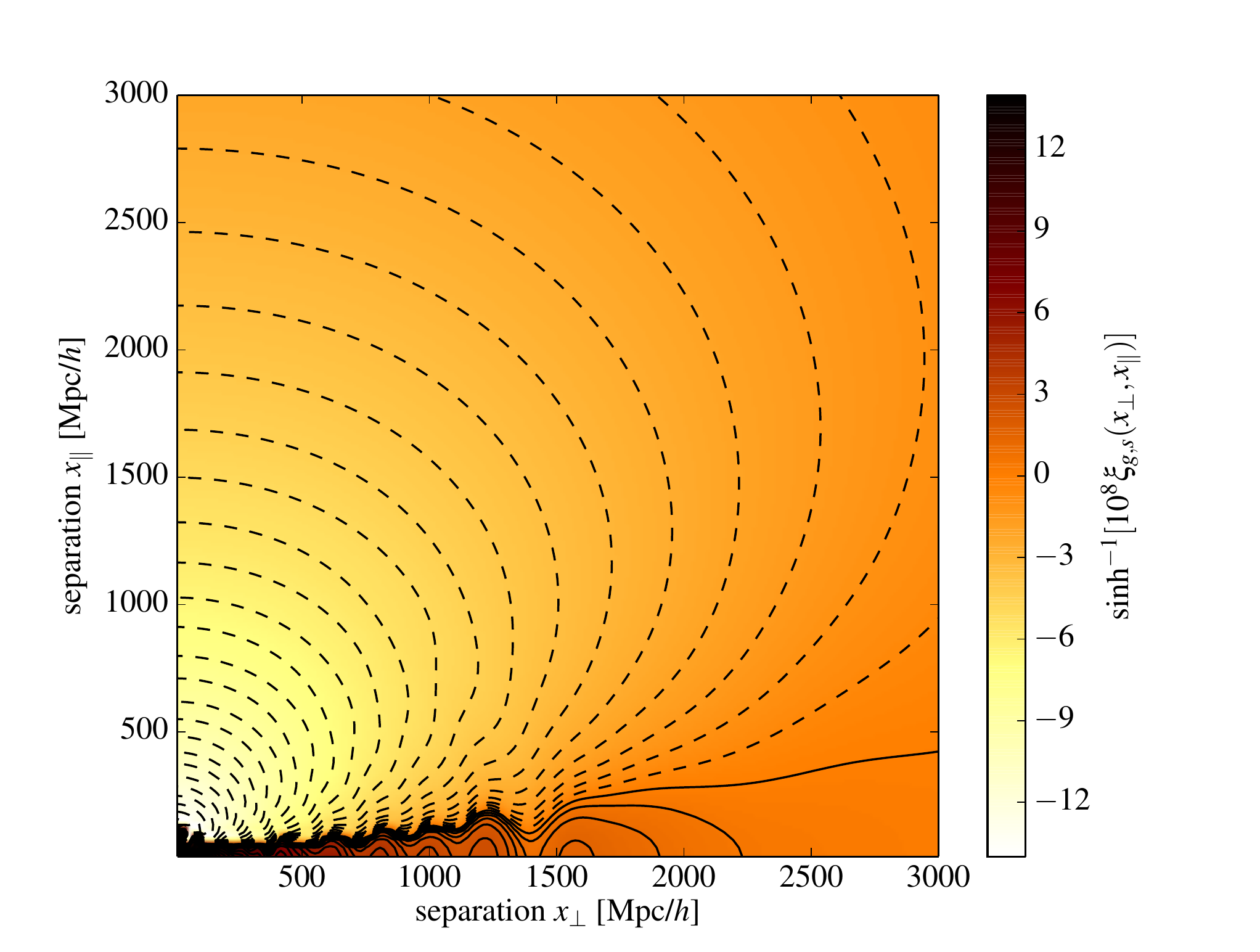}
\caption{
Two-dimensional $(x_\perp,x_\parallel)$ 2PCF from the streaming model for 
larger separations $x<3000~[\mathrm{Mpc}/h]$.
The color scheme and contour lines are for sinh$^{-1}(10^8f)$ where $f$ 
is the function to be shown, indicated in the label of colorbar.
\textit{Top left:} isotropic, real space 2PCF
\textit{Top right:} Correction due to the mean displacement,
$\Delta \xi_{g,s}^{\left<y\right>}$
\textit{Bottom Left:} Correction due to the second moment of the displacement,
$\Delta\xi_{g,s}^{\left<y^2\right>}$
\textit{Bottom right:}
Redshift-space 2PCF from linear Kaiser expression, \refeq{xigs}.
Note that we have not included the wide-angle effect
\citep{szalay/matsubara/landy:1998}, which must be included when the distance
from the observer to the central galaxy is comparable to the perpendicular 
separation $x_\perp$.}
\label{fig:xi2d_streaming_larger}
\end{figure*}
The redshift-space distortion due to the scale-dependent variance of the 
line-of-sight velocity field can be best seen in the streaming model 
\citep{Peebles:1980book,Fisher:1993ye,Fisher:1994ks,Scoccimarro:2004tg}.
Here, pair conservation relates the the redshift-space
2PCF to the real-space 2PCF via coordinate mapping as 
\be
1 + \xi_{g,s}(s_\perp,s_\parallel)
=
\int_{-\infty}^\infty dy 
\left[1 + \xi_{g}(x)\right]
f\left(y;\bfx\right).
\label{eq:streaming}
\ee
Here, $y = s_\parallel-x_\parallel = \bfv_{12}\cdot\nhat/(aH)$ 
is again the displacement ($\bfv_{12}$ is pair-weighted relative velocity,
that we call pairwise velocity for short)
between real space and redshift space, 
and $f(y;\bfx)$ is the line-of-sight directional pairwise velocity 
distribution function, the probability distribution 
of the displacement $y$ at given separation $\bfx = (x_\perp,x_\parallel)$.
As the displacement $y$ is small in linear theory,
we can expand the right hand side of \refeq{streaming} around the position in 
redshift space \citep{Fisher:1994ks,Scoccimarro:2004tg}, and 
\refeq{streaming} becomes
\ba
&1 + \xi_{g,s}(s_\perp,s_\parallel)
\vs
=&
\left[1 + \xi_{g}(s) \right]
\int_{-\infty}^\infty dy 
\biggl[
f\left(y;s_\perp,s_\parallel\right)
+
(x_\parallel - s_\parallel)
\frac{d}{ds_\parallel} f\left(y;s_\perp,s_\parallel\right)
\vs
&+
\frac12 (x_\parallel - s_\parallel)^2
\frac{d^2}{ds_\parallel^2} f\left(y;s_\perp,s_\parallel\right)
\biggl]
+
\mathcal{O}(\xi^2)
\vs
=&
1+ \xi_g(x) 
- 
\frac{d}{dx_\parallel} \left<y\right>
+
\frac12\frac{d^2}{dx_\parallel^2} \left<y^2\right>
+
\mathcal{O}(\xi^2),
\label{eq:xis_streaming_result}
\ea
to linear order in $\xi_g(x)$.
Note that we pull $1+\xi_g(x)=1+\xi_g(s)+\xi_g'(s)(x-s) + \mathcal{O}(\xi^3)
\simeq 1+\xi_g(s)$ outside the integral.
The leading contributions to $\left<y\right>$ and $\left<y^2\right>$ 
are both linear in $\xi(x)$ (see \refeq{ymean} and \refeq{yvar} below for
the explicit expression).
Again, \refeq{xis_streaming_result} tells us that what is relevant for
the linear redshift-space distortion is the line-of-sight
variation of the mean and the variance of the displacement.

We calculate the mean and the variance of the displacement in 
linear theory as follows.  First, we calculate the displacement
$y\equiv \bfv_{12}\cdot\nhat/(aH)$  from the pairwise velocity,
\be
\bfv_{12}(\bfx)
=
\left(\bfv(\bfx_1)-\bfv(\bfx_0)\right)
\left(1+\delta_g(\bfx_1)\right)
\left(1+\delta_{g}(\bfx_0)\right),
\ee
whose leading-order expectation value is given by
\ba
\left< \bfv_{12}(\bfx) \right>
=
\left<\bfv(\bfx_1)\delta_{g}(\bfx_0)\right>
-
\left<\bfv(\bfx_0)\delta_{g}(\bfx_1)\right> + \mathcal{O}(\xi^2).
\ea
Statistical homogeneity guarantees 
$\left<\bfv(\bfx_0)\delta_{g}(\bfx_0)\right> = 
\left<\bfv(\bfx_1)\delta_g(\bfx_1)\right>=0$.
We calculate the velocity field from the Fourier transform,
\be
\bfv(\bfx_0)
= aHf
\int\frac{d^3k}{(2\pi)^3}\frac{i\khat}{k}\delta_m(\bfk)e^{i\bfk\cdot\bfx_0},
\ee
of the continuity equation [\refeq{continuity}].  Then
\ba
\left<\bfv_{12}(\bfx)\right>
=
2\frac{aHf}{b_g}
\int\frac{d^3k}{(2\pi)^3}\frac{i\khat}{k}
P_g(k)
e^{i\bfk\cdot\bfx}
=
- 2\frac{aHf}{b_g}\bfx\xi^1_1(x).
\label{eq:pairwisev}
\ea
Here, we use the identity,
\be
\int \frac{d\Omega_k}{4\pi}\khat e^{i\khat\cdot\bfr} = i \rhat j_1(r).
\ee
Note that the velocity field is radial everywhere, a consequence
of statistical isotropy.  The mean displacement,
\be
\left<y\right> 
=
\frac{1}{aH}\left<\bfv_{12,\parallel}(\bfx)\right>
=
-
2\beta x_\parallel \xi_1^1(x),
\label{eq:ymean}
\ee
here coincides with the result from the heuristic
arguments we presented in \refsec{xi_anisotropy_heuristic}. The
reason behind this correspondence is the pair-conservation
equation \citep{Peebles:1980book},
\be
\frac{\d(1+\xi_g(x))}{\d t} + \frac1a\nabla_{\bfx}\cdot \left<\bfv_{12}(\bfx)\right> = 0,
\ee
which enable us to interpret the pairwise velocity field as the velocity
field associated with radial number density given by 
$n_g(x)\propto 1 + \xi_g(x)$.
Therefore, the heuristic argument in \refsec{xi_anisotropy_heuristic}
correctly captures the redshift-space distortion due to the
mean pairwise velocity $\left<\bfv_{12}\right>$.  We show the
redshift-space 2PCF including this effect of mean displacement
in the top left panel of \reffig{xi2d_streaming}, which is the
same as the bottom right panel of \reffig{xi2d_heuristic}.

Now consider the correction due to the spatial variation of 
the second-order moment. The second moment $\left<y^2\right>$ 
of the displacement is given by the square of the line-of-sight
component of the pairwise velocity,
\ba
\left<
y^2
\right> 
=&
\frac{1}{a^2H^2}\left<\bfv_{12,\parallel}^2(\bfx)\right>
\vs
=&
2\beta^2 
\nhat^i\nhat^j
\int\frac{d^3k}{(2\pi)^3}
\frac{\khat^i\khat^j}{k^2}
P_g(k)
\left[
1 - e^{i\bfk\cdot\bfx}
\right]
\vs
=&
2
\left[
\sigma_v^2 - \beta^2 \int \frac{k^2dk}{2\pi^2}\frac{P_g(kx)}{k^2}
\left(
\frac{j_1(kx)}{kx} - \mu^2j_2(kx)
\right)
\right]
\vs
=& 2
\left[
\sigma_v^2 - \beta^2 x^2 \left(\xi_1^3(x) - \mu^2 \xi_2^2(x) \right)
\right],
\label{eq:yvar}
\ea
where we define the one-dimensional velocity dispersion as
\be
\sigma_v^2 \equiv \frac{\beta^2}{3}\int \frac{dk}{2\pi^2} P_g(k).
\ee
The anisotropies due to the second moment of the displacement 
are determined by $\xi_1^3(x)$ and $\xi_2^2(x)$.  The top-right
panel of \reffig{xi2d_streaming} shows anisotropic contours for
the second-order moment $\left<y^2\right> - 2\sigma_v^2$ of the
displacement field in the two-dimensional $(x_\perp,x_\parallel)$ plane.
As both $\xi_1^3(x)$ and $\xi_2^2(x)$ decrease for large separations,
$\left<y^2\right>$ asymptotes to $\left<y^2\right>\to 2\sigma_v^2$ 
for large separations $x\to \infty$, but it apparently shows 
scale-dependent anisotropies on smaller separations.
On small scales ($x\lesssim100~\mathrm{Mpc}/h$), the
second-order moment becomes smaller than $2\sigma_v^2$ as
$\xi_2^2(x) < \xi_1^3(x)$.  For separations
$x\gtrsim100~\mathrm{Mpc}/h$, however,
$\xi_1^3(x)<\xi_2^2(x)$, and therefore
$\left<y^2\right>>2\sigma_v^2$ for the direction parallel to the
line of sight.  The velocity-dispersion contribution to the
redshift-space 2PCF is given by the second derivative of the second
moment,
\ba
\Delta \xi_{g,s}^{\left<y^2\right>}
\equiv& \frac12\frac{\d^2}{\d x_\parallel^2}\left<y^2\right>
\vs
=&
\beta^2\left[
\frac15 \xi_0^0(x) 
- \frac47\xi_2^0(x)\mathcal{P}_2(\mu)
+ \frac{8}{35}\xi_4^0(x)\mathcal{P}_4(\mu)
\right].
\ea
We show this correction in the bottom left panel of 
\reffig{xi2d_streaming}. Although smaller than the mean
displacement, the basic structure of the anisotropies is the
same: it enhances the 2PCF along the perpendicular direction, and
suppresses it along the parallel direction.

Finally, we show the redshift-space 2PCF including both the mean and the 
second-moment effects of the displacement in the bottom right panel of 
\reffig{xi2d_streaming}. As the mean displacement dominates over the 
second moment, the overall structure of the anisotropies is the same as 
in the top left panel of the same Figure. Including the
spatial variation of the second moment, however, slightly deepens
the negative valley along the line of sight and increases the
clustering amplitude along the parallel direction.

The anisotropic structure in the redshift-space 2PCF persists on larger 
scales as well. The projection of the mean displacement enhances the 2PCF 
along the perpendicular direction and suppresses it along the parallel
direction.  For large separations, however, the
parallel-direction suppression from the mean displacement is
somewhat reduced by the positive contribution from the variance
(although the net effect is still suppression along the parallel
direction).  In \reffig{xi2d_streaming_larger}, we show the
redshift-space 2PCF (bottom right) for larger separations
($x<3000~\mathrm{Mpc}/h$) along  with the real-space 2PCF (top
left),  the mean-displacement correction (top right), and 
the second-moment correction (bottom left). 
In order to compensate the small 2PCF amplitude at these larger separations,
we show $\sinh^{-1}(10^8\xi_g)$ here.
We caution the readers here that the contour plot in 
\reffig{xi2d_streaming_larger} is correct only when the distance from the
observer to the galaxies is much larger compared to the radial separation
$x_\perp$. Otherwise, the plane-parallel approximation is broken down, and 
wide-angle effect \citep{szalay/matsubara/landy:1998} must be included.
We will present the contour plot including the wide-angle effect elsewhere
\citep{jeong/etal:inprep}.

The upper-most solid curve in the bottom right panel shows the 
zero-crossing trajectory. Unlike the real-space 2PCF (hence the
monopole 2PCF) case that the zero-crossing separation can be a
proxy for the ratio between matter and radiation energy density 
(through matter-radiation equality redshift) \citep{prada/etal:2011}, the
zero-crossing trajectory in the two-dimensional redshift-space
2PCF may also depend on the linear-theory growth factor $f$  
and the galaxy bias $b_g$ through \refeq{xigs}.
We defer discussion about the cosmological information contained in 
the zero-crossing trajectory as well as its robustness to future work.
We stress, however, here that a study of the zero-crossing trajectory on
large scales must include the full description of the redshift-space 
distortion beyond the plane-parallel approximation we employ here.

\section{Redshift-space distortion of  the BAO}
\label{sec:BAO}
\begin{figure*}
\centering
\includegraphics[width=0.498\textwidth]{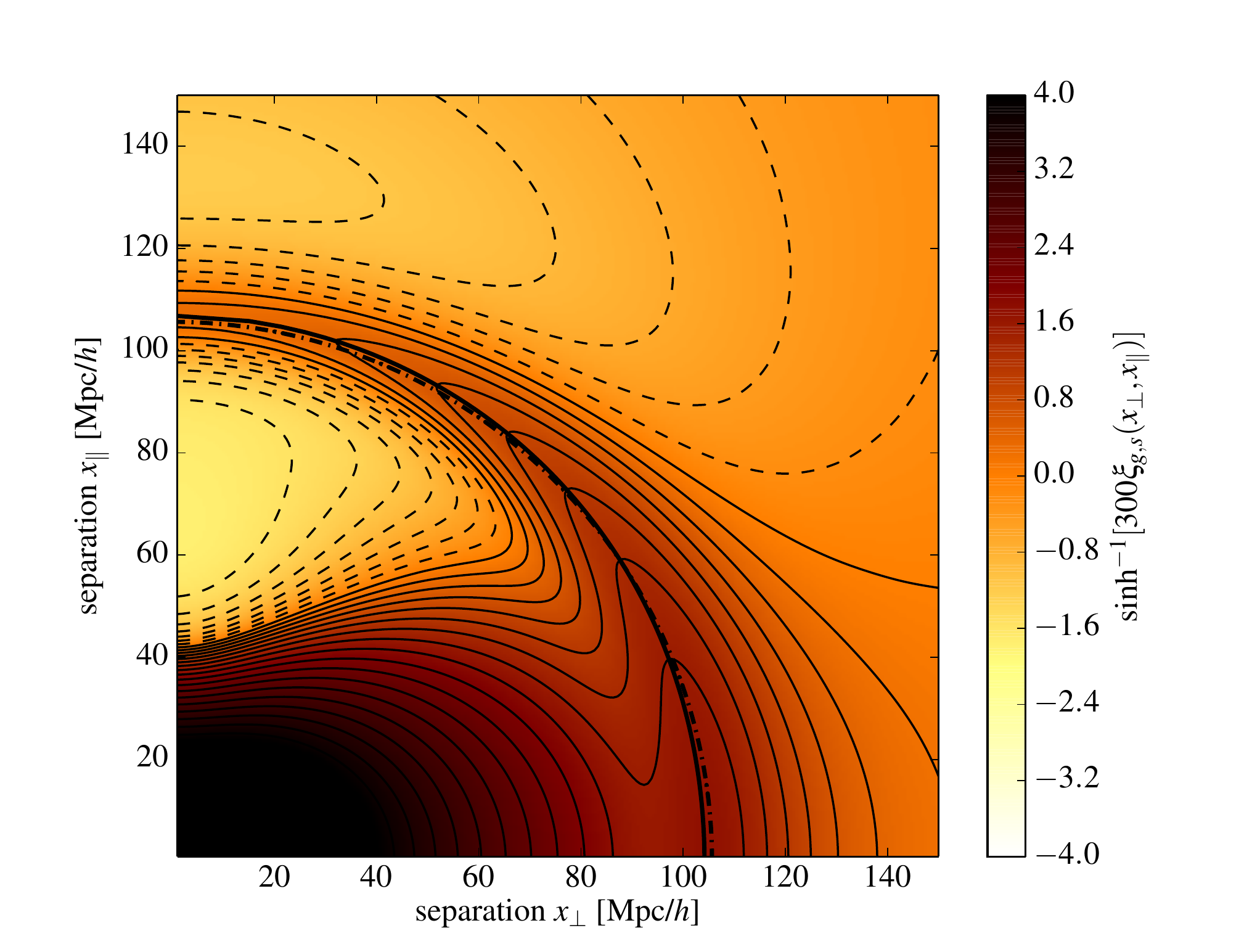}
\includegraphics[width=0.498\textwidth]{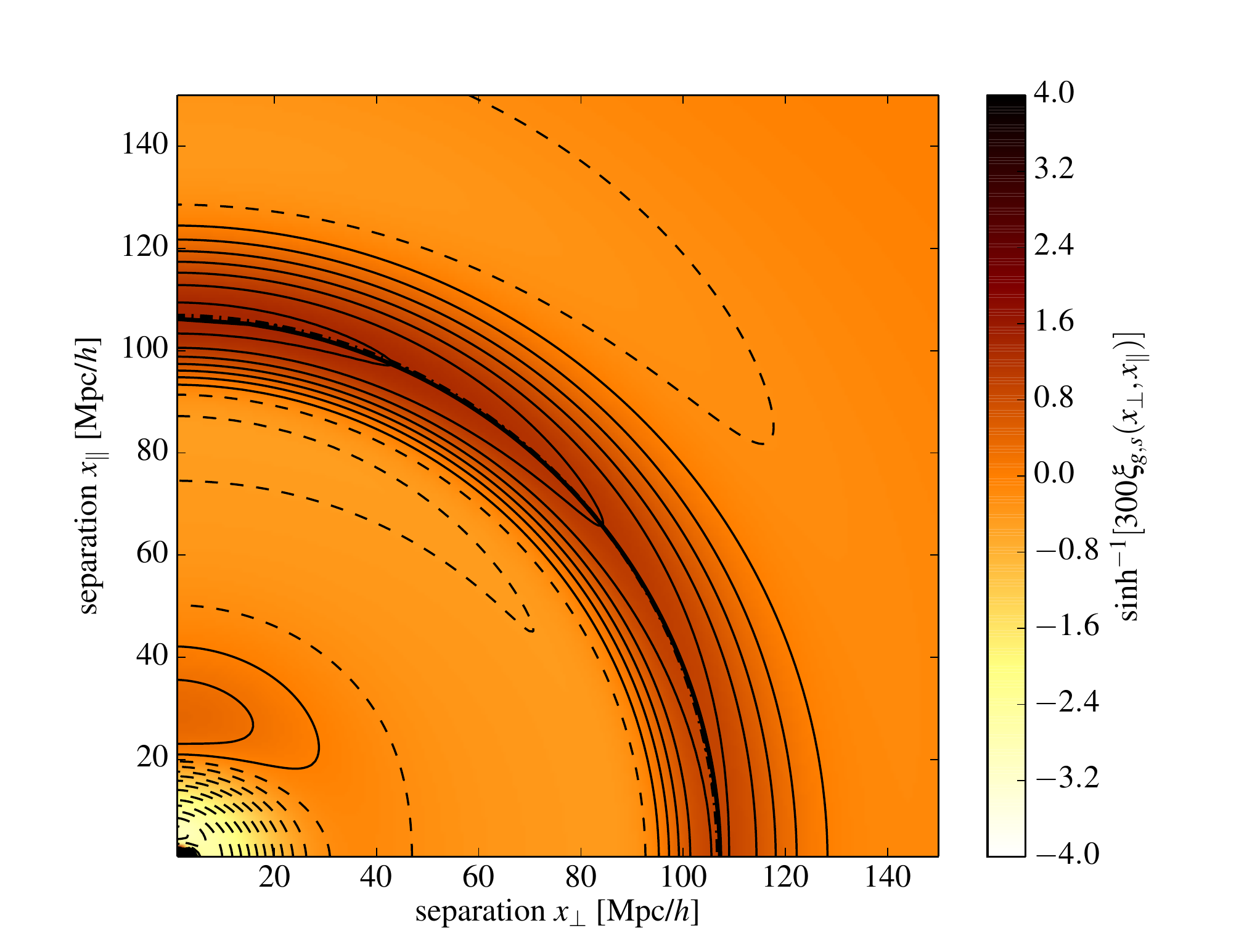}
\caption{
Location of BAO peaks in real space (dotted magenta lines) and 
redshift space (solid magenta lines) in the two-dimensional 
$(x_\perp,x_\parallel)$ plane.
\textit{Left:} BAO estimated from peak locations of the redshift-space 2PCF.
\textit{Right:} BAO estimated from peak locations of 
the redshift-space 2PCF, subtracting the anisotropic `background' due to the 
redshift-space distortion of the no-wiggle 2PCF $\xi_g^{\mathrm{nw}}(x)$.
We calculate the no-wiggle 2PCF from \refeq{xigs} 
with the no-wiggle power spectrum in \citet{Eisenstein:1997ik}.
}
\label{fig:xi2d_BAO}
\end{figure*}
\begin{figure*}
\centering
\includegraphics[width=0.498\textwidth]{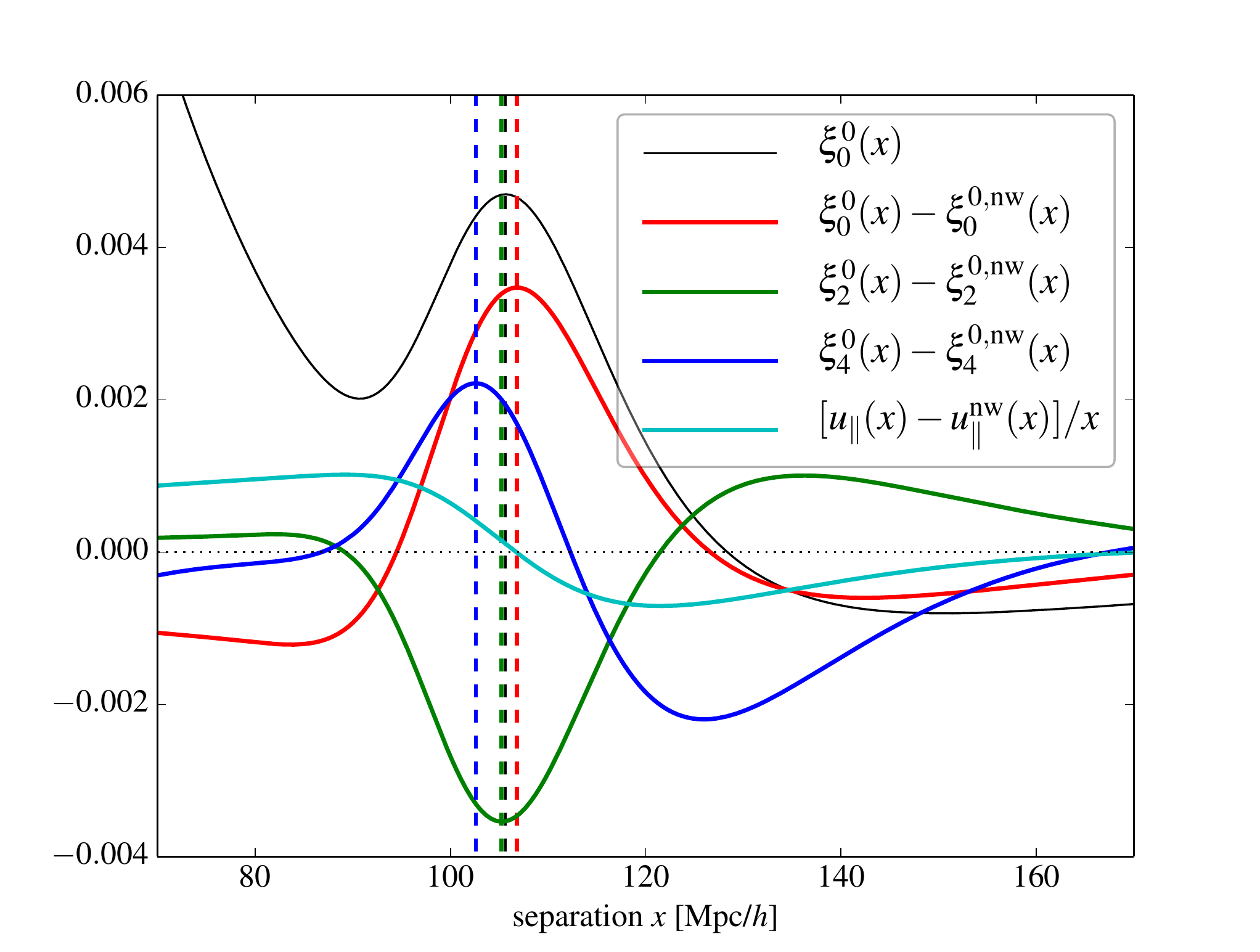}
\includegraphics[width=0.498\textwidth]{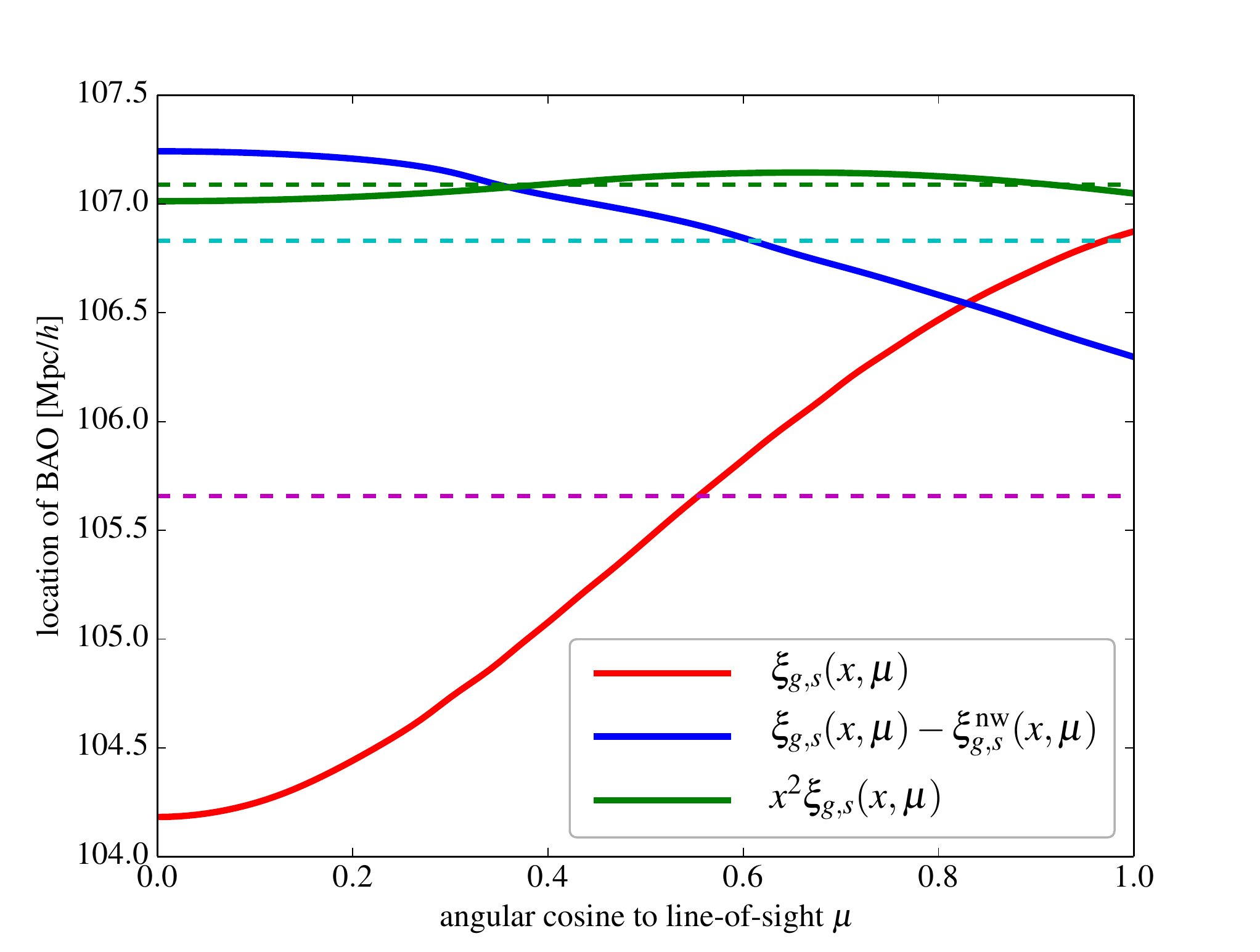}
\caption{
\textit{Left:}
Location of BAO peaks in the real-space galaxy 2PCF $\xi_0^0(x)$ 
(\textit{black}) as well as the wiggly parts of 
$\xi_0^0(x)$, $\xi_2^0(x)$, and $\xi_4^0(x)$ which are calculated by
subtracting the corresponding no-wiggle 2PCF $\xi_n^{m,\mathrm{nw}}(x)$. 
The no-wiggle $\xi_n^{m,\mathrm{nw}}(x)$ are calculated from \refeq{xigs}
with the no-wiggle power spectrum in \citet{Eisenstein:1997ik}.
Vertical dotted lines show the peak locations:
$r_\mathrm{BAO} = 105.6~\mathrm{Mpc}/h$,
$106.8~\mathrm{Mpc}/h$,
$105.2~\mathrm{Mpc}/h$, and 
$102.6~\mathrm{Mpc}/h$ for, respectively,
$\xi_0^0$ (black dotted),
$\xi_0^0-\xi_0^{0,\mathrm{nw}}$ (red dotted),
$\xi_2^0-\xi_2^{0,\mathrm{nw}}$ (green dotted), and
$\xi_4^0-\xi_4^{0,\mathrm{nw}}$ (blue dotted).
The cyan line shows the amplitude of the `local' peculiar-velocity field
re-scaled by $x_\parallel$: $[u(x) - u^{\rm nw}(x)]/x_\parallel$.
The local velocity field vanishes around the BAO peak of 
$\xi_0^0(x) - \xi_0^{0,\rm nw}(x)$ (red solid line).
\textit{Right:}
Location of BAO peaks as a function of the angular cosine $\mu$ to the LoS.
Note that the BAO peak location depends on the method to measure the peak:
red solid line and magenta dotted line are redshift-space and real space
BAO peak location measured from the peak position of 
$\xi_{g,s}(x,\mu)$ for fixed $\mu$, while blue solid line and
the cyan dotted 
line are the BAO peak location measured from the peak position of 
$\xi_{g,s}(x,\mu)-\xi_{g,s}^{\mathrm{nw}}(x,\mu)$, where 
$\xi_{g,s}^{\mathrm{nw}}(x,\mu)$ is the redshift-space 2PCF
without the BAO bump. Green dashed line and the solid line are the same
for $x^2\xi_{g,s}(x,\mu)$.
}
\label{fig:xBAO}
\end{figure*}
In the previous Section, we have shown that the anisotropies in
the redshift-space 2PCF strongly depend on scale, and the
linear-theory expression for the scale dependence can be
explained by the line-of-sight variation in the first (mean) and
the second moment of the displacement.  Then, how do these
anisotropies affect the baryon acoustic oscillation (BAO)?

In both \reffig{xi2d_heuristic} and \reffig{xi2d_streaming}, we
show the location of the real-space BAO peak ($x\approx
106~\mathrm{Mpc}/h$) as an isotropic thick solid line. By
comparing the real-space 2PCF (top left panel of
\reffig{xi2d_heuristic}) to the redshift-space 2PCF (bottom
right panel of \reffig{xi2d_streaming}), we first notice that
the the amplitude of the BAO bump follows the general structure
of the redshift-space distortion: the BAO bump is enhanced along
the perpendicular direction and suppressed along the parallel direction.
We also find that the location of the local BAO `peak' and the width of the 
BAO bump both vary, and the variation is increasingly apparent as 
we approach the line-of-sight direction.  In this Section, we
shall quantify the anisotropies in the BAO amplitude as well as
shift of the BAO peak location in redshift space.

First, we study the anisotropic shift of the BAO peak in the
redshift-space 2PCF.  The simplest way to define the BAO peak is
by finding a local maximum around the radial separation of $x\approx
105~\mathrm{Mpc}/h$, which is shown in the left panel of
\reffig{xi2d_BAO} as a thick solid line. In order to facilitate
the comparison, we also show the quarter circle with radius of 
$d_{\rm BAO}$ as a thick dot-dashed line. 
The anisotropic shift of the BAO
peaks is apparent in this plot: BAO peaks move away from the
real-space location along the line-of-sight direction, and move
toward the center along the perpendicular direction.  We show
the resulting BAO peak location in the right panel of
\reffig{xBAO} (red solid line) as a function of the angular
cosine $\mu$ between the line of sight and the
separation.  The shift of BAO peaks due to the redshift-space
distortion is biggest at each end of the parallel and
perpendicular direction, and the maximum shift is about
$\Delta_{\rm BAO} \approx 1.5~\mathrm{Mpc}/h$ compared to the
real space BAO position (dashed magenta line).

While the shift is apparent for the BAO peak defined as a local maximum
in the two-dimensional 2PCF, some fraction of this shift may arise from the 
redshift-space distortion associated with the broadband tilt of
the 2PCF.
In fact, the direction that the BAO peak shifts in \reffig{xi2d_BAO} 
coincides with the direction to which the 2PCF 
correction due to redshift-space distortions increases.  As discussed in
\refsec{xigs_model}, the redshift-space distortion suppresses
the 2PCF along the line of sight and enhances it along the
perpendicular direction.  Also, the amount of suppression and
enhancement is larger for the smaller separation. Adding these
anisotropies in slope around the BAO naturally moves the peak to
larger (smaller) separations along the line-of-sight
(perpendicular) direction as we observe in the left panel of
\reffig{xi2d_BAO}.

We can remove this shift of the BAO peak from the overall shape of
the 2PCF by subtracting the broadband shape of the
redshift-space 2PCF. To do so, we estimate the BAO-less
broadband redshift space 2PCF from the linear redshift-space
2PCF (\refeq{xigs}) with the no-wiggle power spectrum in 
\citet{Eisenstein:1997ik}. The right panel of \reffig{xi2d_BAO} shows 
the resulting contour plot which highlights the BAO bump in 
the redshift-space 2PCF. Again, two thick lines show the BAO 
in real space (dot-dashed) and in redshift space (solid). 
The measured peak locations from the two-dimensional contour
plot are shown in the right panel of \reffig{xBAO} with the
dashed cyan line (real space) and the solid blue line (redshift
space) as a function of $\mu$.  When subtracting the overall
redshift-space distortion signature in the broadband 2PCF, the
measured BAO peaks in redshift space are less anisotropic than
before, although it is still not a perfect circle.  As before,
the shift of the BAO peak is greatest at each end: the parallel   
($\mu=1$) and the perpendicular ($\mu=0$) directions with 
the maximum shift of about a third of the previous case 
($\Delta_{\rm BAO}\approx0.5~\mathrm{Mpc}/h$).

The shift of the BAO peak in this case can also be seen in the left panel of
\reffig{xBAO}, where we plot the real-space 2PCF ($\xi_0^0(x)$,
black line) along with the {\it BAO components} of $\xi_0^0(x)$ (red
line), $\xi_2^0(x)$ (green line), and $\xi_4^0(x)$ (blue
line). We calculate the BAO component by subtracting  
$\xi_n^m(x)$ from the bump-less $\xi_n^m(x)$ using the no-wiggle power spectrum 
in \citet{Eisenstein:1997ik}. The vertical dashed lines show the
location of peak for the lines of corresponding colors. As can
be seen in the Figure, the locations of the peaks are different for
all cases:
$\xi_0^0 (105.6~\mathrm{Mpc}/h)$,
$\xi_0^0 -\xi_0^{0,\mathrm{nw}} (106.8~\mathrm{Mpc}/h)$,
$\xi_2^0 -\xi_2^{0,\mathrm{nw}} (105.2~\mathrm{Mpc}/h)$, and 
$\xi_4^0 -\xi_4^{0,\mathrm{nw}} (102.6~\mathrm{Mpc}/h)$.
Since the bump in the redshift-space 2PCF is, after subtracting
the broadband shape, given by a linear combination of the
latter three functions with coefficients given by \refeq{xigs},
the locations of the BAO peaks must show anisotropies.

\begin{figure*}
\centering
\includegraphics[width=0.498\textwidth]{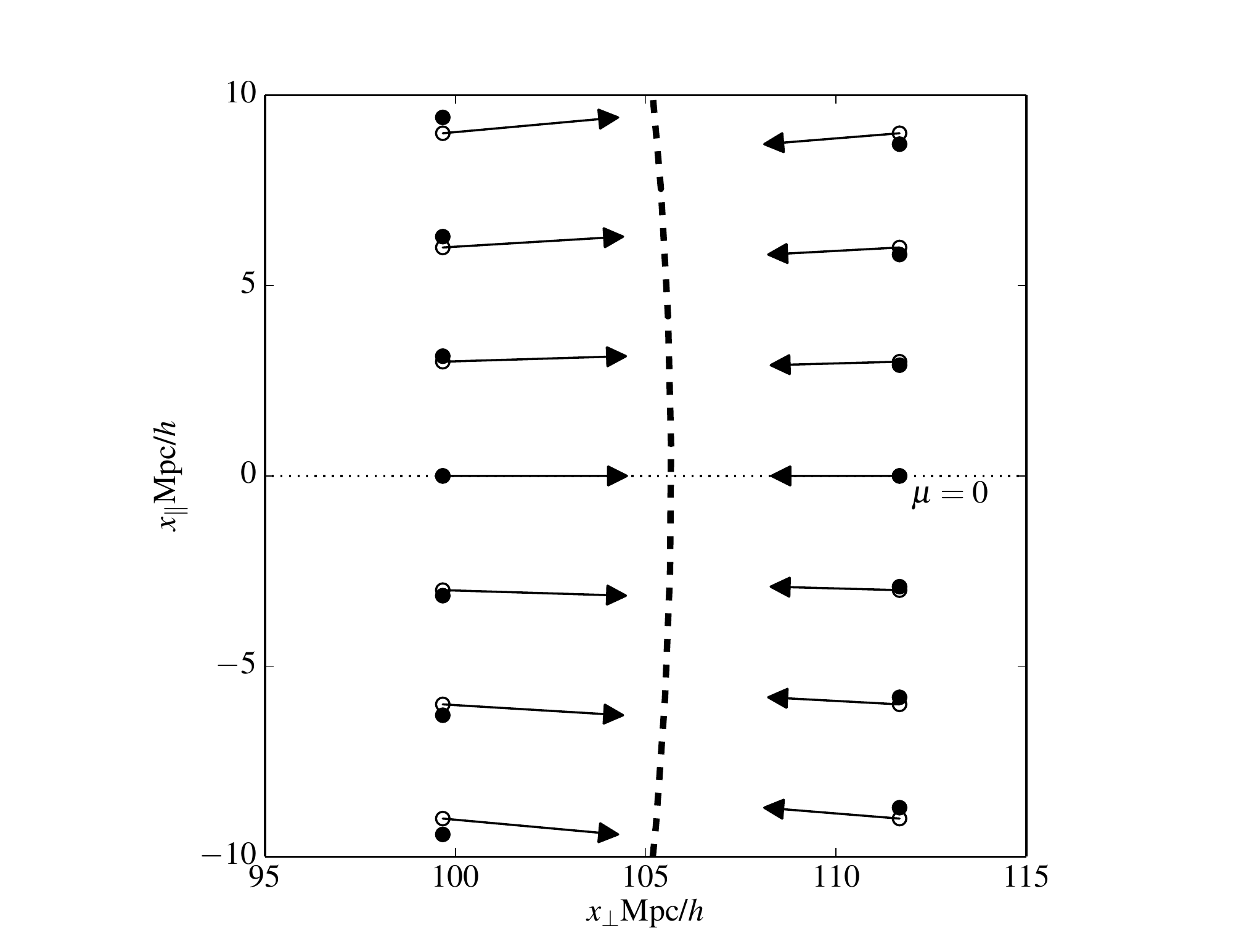}
\includegraphics[width=0.498\textwidth]{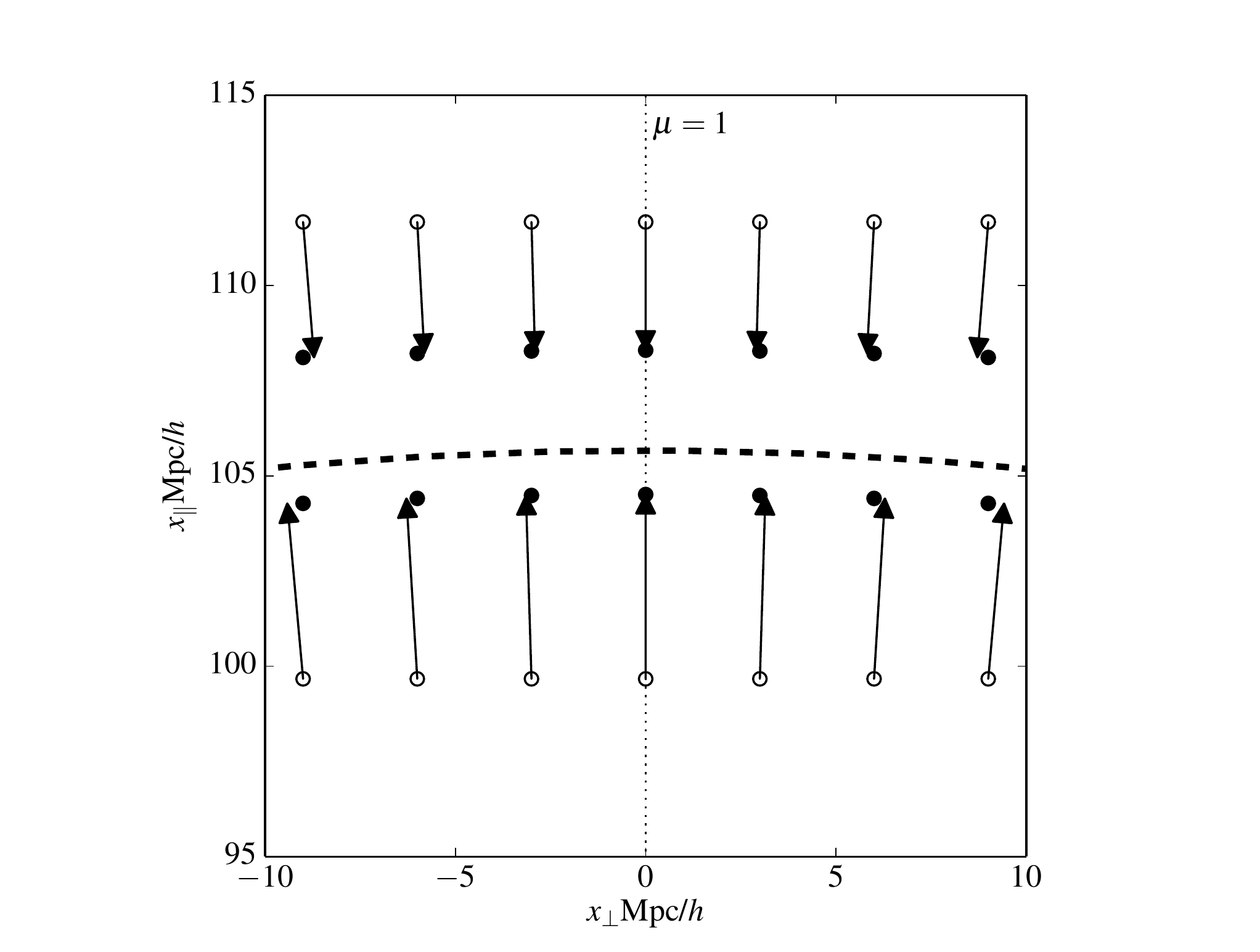}
\caption{
A sketch illustrating the shift of BAO peaks after subtracting the 
contribution from broadband redshift-space distortions.
The left panel shows the effect around the perpendicular axis ($\mu=0$), and
the right panel shows it around the parallel axis ($\mu=1$).
For both cases, the peculiar-velocity field is toward the BAO peak
location, represented by the thick dashed line at 
$d_{\rm BAO}\simeq 105~{\rm Mpc}/h$.
The empty circles are real-space positions at constant 
$x_\perp$ (left) and $x_\parallel$ (right), while the filled circles are 
the corresponding redshift-space positions. To illustrate, we 
amplify the peculiar velocity by a factor of 100.
For $\mu=0$ ($\mu=1$), redshift-space distortions increase the number 
density at larger (smaller) separation; thus, the BAO peaks shift toward
larger (smaller) separation.
}
\label{fig:sketch_BAO}
\end{figure*}
In \reffig{xBAO}, we notice that the estimated BAO peaks after
removing the broadband shape shift to the opposite direction
than the case before the removal.  Why are BAO peaks shifted in
this way?  To understand that, let us go back to the heuristic
galaxy distribution that we have considered in
\refsec{xi_anisotropy_heuristic}.  In linear theory, the radial
density field (set by 2PCF) is amplified by the linear growth
factor, and the peculiar-velocity field is set to accommodate
the growth.  The overall infalling velocity field feeds the
inner over-densities, and its line-of-sight projection generates
the redshift-space distortion of the broadband 2PCF.
Furthermore, the BAO bump in the radial profile forms a
spherical shell of over-density around $x\approx
105~[\mathrm{Mpc}/h]$.  This then generates additional velocity
structure locally converging onto the peak so that the peak
maintains linear growth.  It is this additional local velocity
field that shifts the BAO peaks even after subtracting the
broadband redshift-space distortion.
The local velocity field vanishes at the peak and increases its 
amplitude as it moves away from the peak at the vicinity of the BAO peak.
Along the perpendicular direction
($\mu=0$), the sign of this additional redshift-space distortion 
follows the sign of the additional infalling velocity field 
(see \refeq{yterm}): it increases the 2PCF for
$x\gtrsim106~\mathrm{Mpc}/h$ (where $y>0$) and decreases for
$x\lesssim106~\mathrm{Mpc}/h$ (where $y<0$) so that the BAO peak shifts
toward the larger separation.  The BAO peak also shifts along
the parallel direction ($\mu=1$), even though the
peculiar-velocity field (cyan line in the left panel of
\reffig{xBAO}) vanishes at the peak location of
$\xi_0^0(x)-\xi_0^{0,{\rm nw}}(x)$.  The asymmetry of the
local peculiar-velocity field around the BAO makes the
derivative of the local velocity peak at smaller $x$, 
around the peak of $\xi_2^0-\xi_2^{0,\mathrm{nw}}$ (green line in the 
left panel of \reffig{xBAO}).  
We illustrate this point in the sketch shown in \reffig{sketch_BAO}.
The shift due to the second
derivative of the velocity dispersion, which is not included in the
heuristic argument, further shifts the BAO to the same
direction. However, the amplitude of the shift is dominated by the
mean velocity. 

Finally, we find  that the BAO peak location detected from the volume 
weighted 2PCF, $x^2\xi(x,\mu)$, is significantly more stable than the
previous two methods. The BAO location measured from $x^2\xi(x)$ 
(real space 2PCF) and from $x^2\xi_g(x,\mu)$ (redshift space 2PCF) 
are shown as, respectively, dashed and solid green lines in \reffig{xBAO}.
In this case, we only observe sub-percent level of angular variation
in the BAO peak location, because the additional factor of $x^2$
cancels out the BAO shift due to the radial peculiar velocity 
$d\left<y\right>/dx_\parallel$ term in \refeq{xis_streaming_result} and the
shift is mostly due to the spatial variation of the line-of-sight directional
velocity dispersion.
Specifically, the solution $x=x_0$
of the equation $[x^2\xi(x)]'=0$ that defines the peak of BAO 
also satisfies 
\be
\frac{d}{dx}\left[x^2 \frac{d\left<y\right>}{dx_\parallel}\right]
\propto
\frac{d}{dx}\left[
x^2
\left(
\xi(x) - 
\frac{1}{x^3}\int_0^x dy y^2 \xi(y)\right)
\right]=0,
\ee
which keeps the BAO peak location constant when adding the effect of 
infalling bulk velocity.

In addition to the shift of the BAO peak, we also observe that after 
removing the broadband redshift-space distortion, the BAO bump
becomes higher and sharper along the line of sight.  We can
understand this feature, again, from the heuristic model in
\refsec{xi_anisotropy_heuristic}: the local velocity field moves
nearby galaxies close to the BAO peak, so that the BAO peak in
redshift space is sharper towards the line of sight.  This
interesting feature was first observed by \citet{Tian:2010wx}
when they identify the BAO with a matching filter.

\section{Conclusion}
\label{sec:conclusion}
We present a detailed explanation for the anisotropies in the 
linear-theory galaxy 2PCF in redshift space. Because they are
related by a non-local operation---namely, Fourier
transformation---the  anisotropies in the galaxy 2PCF are
manifest quite differently from the anisotropies in the galaxy
power spectrum. For the galaxy power spectrum, the angular
dependence can be well separated from the scale dependence, and
the redshift-space distortion preserves the power in the
perpendicular directional.  On the other hand, the angular
dependence appears differently in different scales for the 2PCF,
and redshift-space distortions affect the 2PCF along all directions.

To develop intuition for the anisotropic 2PCF, we consider a
typical mass and galaxy distribution around a central galaxy and
the associated peculiar-velocity field (via the continuity equation)
that maintains the linear-theory growth of the large-scale
density fluctuation.  The peculiar-velocity field that feeds the
growth of the over-densities around the central galaxy is
radial and decreases for large separations.  Then, the
redshift-space distortion, proportional to the parallel
derivative of the line-of-sight component of the
peculiar-velocity field, enhances the galaxy 2PCF along the
perpendicular direction and toward small separations.
In addition to the mean peculiar velocity, we show with the 
streaming model that the second derivative of the second
velocity moment must be included to get the $f^2$ terms in the
linear redshift-space distortion.

We then explored with this model the anisotropic shift of BAO in
the  two-dimensional redshift space.  Although the BAO peaks
shift in the two-dimensional 2PCF, when averaging over 
angles, the BAO peak in the monopole 2PCF is still at the
position of the real-space BAO, as the monopole 2PCF (the
$\mu$-independent terms in \refeq{xigs}) is proportional to
$\xi_0^0(x)$.  Previous studies also showed that the location of
the monopole BAO is robust even in the presence of
non-linear redshift-space distortions.  That we discover a
percent-level shift of the BAO peak in redshift space,
even in linear theory, however, suggests that one has to fully
understand the angular-dependence in the redshift-space 2PCF
including full non-linearities before using the full
two-dimensional measurement  of BAO for, e.g., the
Alcock-Paczynski test. If naively assuming that BAO peaks are
isotropic, the distance measurement using BAO will be
systematically shifted.

As we have demonstrated in \refsec{BAO}, the shift of the BAO
peak depends on the method to extract the BAO peak. There are
several methods to subtract the broadband shape of 2PCF and
identifying BAO, and the amount of shift may depend on the
method of measuring BAO peak location. For example,
\citet{Tian:2010wx} have used a Mexican-hat wavelet
transformation to identify the location of BAO peaks and find
that the shift of BAO is within statistical uncertainties of
their Gaussian simulation. As the statistical uncertainties in
that work are $\simeq1~\mathrm{Mpc}/h$, in order to
accommodate the need for accurate measurement of BAO in future
galaxy surveys, more detailed studies, which we leave for future
work, are in order.
One can also reduce the shift of BAO peak by using the non-linear 
transform such as the log-density transformation \citep{mccullagh/etal:2013},
and future extension of this method must include the analysis 
for the biased tracers such as galaxies.

Finally, the galaxy 2PCF on scales larger than BAO can also be
used to measure the logarithmic growth index $f$ and
geometrical quantities like the angular-diameter distance $D_A(z)$
and Hubble expansion rate $H(z)$.  On such large scales, the
density field is linear, but one needs to include the
correction to the usual Kaiser formula due to the curvature of
the sky, radial evolution of cosmological parameters and
galaxy number density, and general-relativistic
corrections.  We shall present the full two-dimensional 2PCF
including all of these effects elsewhere \citep{jeong/etal:inprep}.

\section*{Acknowledgments}
DJ, MK and LD were supported by the John Templeton
Foundation and NSF grant PHY-1214000.  
AS has been supported by the Gordon and Betty Moore Foundation,
and NSF OIA-1124403.

\bibliographystyle{mn2e}
\bibliography{tamred}

\end{document}